\documentclass{article}
\usepackage{eurosym}
\usepackage{algorithm}
\usepackage{amsfonts}
\usepackage{setspace}
\usepackage{wrapfig}
\usepackage{subfig}
\usepackage{graphicx}
\usepackage[margin=1in]{geometry}
\usepackage{float}
\usepackage[toc,page]{appendix}
\usepackage{color}
\usepackage{amsmath}
\usepackage{setspace}
\usepackage{float}
\usepackage{amssymb}

\newtheorem{theorem}{Theorem}
\newtheorem{acknowledgement}[theorem]{Acknowledgement}

\input{tcilatex}
\begin{document}

\title{Old Problems, Classical Methods, New Solutions}
\author{ Alexander Lipton\thanks{%
The Jerusalem School of Business Administration, The Hebrew University of
Jerusalem, Jerusalem, Israel; Connection Science and Engineering,
Massachusetts Institute of Technology, Cambridge, MA, USA; SilaMoney,
Portland, OR, USA; Investimizer, Chicago, IL, USA; E-mail: alexlipt@mit.edu} 
}
\date{}
\maketitle

\begin{abstract}
We use a powerful extension of the classical method of heat potentials,
recently developed by the present author and his collaborators, to solve
several significant problems of financial mathematics. We consider the
following problems in detail: (A) calibrating the default boundary in the
structural default framework to a constant default intensity; (B)
calculating default probability for a representative bank in the mean-field
framework; (C) finding the hitting time probability density of an
Ornstein-Uhlenbeck process. Several other problems, including pricing
American put options and finding optimal mean-reverting trading strategies,
are mentioned in passing. Besides, two nonfinancial applications - the
supercooled Stefan problem and the integrate-and-fire neuroscience problem -
are briefly discussed as well.
\end{abstract}

\section{Introduction\label{Intro}}

The method of heat potentials (MHP) is a highly robust and versatile
approach frequently exploited in mathematical physics; see, e.g., \cite%
{Tikhonov1963, Rubinstein1971, Kartashov2001, Wilson2012} among others. It
is essential in numerous vital fields, such as thermal engineering, nuclear
engineering, and material science.

However, it is not particularly popular in mathematical finance, even though
the first meaningful use case was described by the present author almost
twenty years ago. The specific application was to pricing barrier options
with curvilinear barriers, see \cite{Lipton2001}, Section 12.2.3.

In this document, we demonstrate how a powerful extension of the classical
MHP, recently developed by the present author and his collaborators, can be
used to solve seemingly unrelated problems of applied mathematics in general
and financial mathematics in particular, see \cite{Lipton2018b, Lipton2020a,
Lipton2020b, Lipton2020c, Lipton2020d}.

Specifically, we use the extended method of heat potentials (EMHP) for (A)
calibrating the default boundary for a structural default model with
constant default intensity; (B) finding a semi-analytical solution of the
mean-field problem for a system of interacting banks; (C) developing a
semi-analytical description for the hitting time density for an
Ornstein-Uhlenbeck process. Besides, we demonstrate the efficacy of the EMHP
by considering two nonfinancial applications: (A) the supercooled Stefan
problem; (B) the integrate-and-fire model in neuroscience.

We note in passing that, in addition to the problems discussed in this
document, the EMHP has been successfully used for pricing American put
options and for finding optimal strategies for mean-reverting spread
trading, see \cite{Lipton2020a, Lipton2020b}.

We emphasize that in most cases, the EMHP beats all other known approaches
to the problem in question, and in some instances, for example, for the
boundary calibration problem, it is the only one that can be used
effectively.

\section{Mathematical preliminaries\label{Mathprel}}

In this section, we describe the classical MHP and its beneficial extensions
proposed by the author and his collaborators.

\subsection{The method of heat potentials}

Consider a standard heat equation in a one-sided domain with a moving
boundary $b^{>}\left( t\right) $:

\begin{equation}
\begin{array}{c}
\frac{\partial }{\partial t}E^{>}\left( t,x\right) =\frac{1}{2}\frac{%
\partial ^{2}}{\partial x^{2}}E^{>}\left( t,x\right) ,\ \ \ b^{>}\left(
t\right) \leq x<\infty , \\ 
\\ 
E^{>}\left( 0,x\right) =\varepsilon ^{>}\left( x\right) ,\ \ \ E^{>}\left(
t,b^{>}\left( t\right) \right) =e^{>}\left( t\right) ,\ \ \ E\left(
t,x\rightarrow \infty \right) \rightarrow 0.%
\end{array}
\label{Eq1}
\end{equation}%
Without loss of generality, we can assume that $\varepsilon ^{>}\left(
x\right) =0$; the case of a nonzero initial condition can be solved by
splitting:%
\begin{equation*}
\begin{array}{c}
E^{>}=E\left( t,x\right) +F^{>}\left( t,x\right) ,%
\end{array}%
\end{equation*}%
\begin{equation*}
\begin{array}{c}
E\left( t,x\right) =\int_{b\left( t\right) }^{\infty }H\left( t,x-y\right)
\varepsilon ^{>}\left( y\right) dy,%
\end{array}%
\end{equation*}%
where $H\left( t,x\right) $ is the standard heat kernel,%
\begin{equation*}
\begin{array}{c}
H\left( t,x\right) =\frac{e^{-\frac{x^{2}}{2t}}}{\sqrt{2\pi t}}.%
\end{array}%
\end{equation*}%
Thus, we can restrict ourselves to the case of zero initial condition:%
\begin{equation*}
\begin{array}{c}
\frac{\partial }{\partial t}F^{>}\left( t,x\right) =\frac{1}{2}\frac{%
\partial ^{2}}{\partial x^{2}}F^{>}\left( t,x\right) ,\ \ \ b^{>}\left(
t\right) \leq x<\infty , \\ 
\\ 
F^{>}\left( 0,x\right) =0,\ \ \ F^{>}\left( t,b^{>}\left( t\right) \right)
=f^{>}\left( t\right) ,\ \ \ F\left( t,x\rightarrow \infty \right)
\rightarrow 0,%
\end{array}%
\end{equation*}%
where%
\begin{equation*}
\begin{array}{c}
f^{>}\left( t\right) =e^{>}\left( t\right) -E\left( t,b\left( t\right)
\right) .%
\end{array}%
\end{equation*}%
The MHP allows one to represent $F^{>}\left( t,x\right) $ in the form%
\begin{equation}
\begin{array}{c}
F^{>}\left( t,x\right) =\int_{0}^{t}\frac{\left( x-b^{>}\left( t^{\prime
}\right) \right) \exp \left( -\frac{\left( x-b^{>}\left( t^{\prime }\right)
\right) ^{2}}{2\left( t-t^{\prime }\right) }\right) \nu ^{>}\left( t^{\prime
}\right) }{\sqrt{2\pi \left( t-t^{\prime }\right) ^{3}}}dt^{\prime },%
\end{array}
\label{Eq2}
\end{equation}%
where $\nu ^{>}\left( t\right) $ solves the Volterra equation of the second
kind:%
\begin{equation}
\begin{array}{c}
\nu ^{>}\left( t^{\prime }\right) +\int_{0}^{t}\frac{\Theta ^{>}\left(
t,t^{\prime }\right) \Xi ^{>}\left( t,t^{\prime }\right) \nu ^{>}\left(
t^{\prime }\right) }{\sqrt{2\pi \left( t-t^{\prime }\right) }}dt^{\prime
}=f^{>}\left( t\right) ,%
\end{array}
\label{Eq3}
\end{equation}%
and%
\begin{equation*}
\begin{array}{c}
\ \Theta ^{>}\left( t,t^{\prime }\right) =\frac{b^{>}\left( t\right)
-b^{>}\left( t^{\prime }\right) }{\left( t-t^{\prime }\right) },\ \ \ \Xi
^{>}\left( t,t^{\prime }\right) =e^{-\frac{\left( t-t^{\prime }\right)
\Theta ^{>2}\left( t,t^{\prime }\right) }{2}}, \\ 
\\ 
\Theta ^{>}\left( t,t\right) =\frac{db^{>}\left( t\right) }{dt},\ \ \ \Xi
^{>}\left( t,t\right) =1.%
\end{array}%
\end{equation*}

Similarly, the solution to the problem%
\begin{equation*}
\begin{array}{c}
\frac{\partial }{\partial t}F^{<}\left( t,x\right) =\frac{1}{2}\frac{%
\partial ^{2}}{\partial x^{2}}F^{<}\left( t,x\right) ,\ \ \infty <\ x\leq
b^{<}\left( t\right) , \\ 
\\ 
F^{<}\left( 0,x\right) =0,\ \ \ F^{<}\left( t,x\rightarrow -\infty \right)
\rightarrow 0,\ \ \ F^{<}\left( t,b\left( t\right) \right) =f^{<}\left(
t\right) ,%
\end{array}%
\end{equation*}%
has the form%
\begin{equation*}
\begin{array}{c}
F^{<}\left( t,x\right) =\int_{0}^{t}\frac{\left( x-b^{<}\left( t^{\prime
}\right) \right) \exp \left( -\frac{\left( x-b^{<}\left( t^{\prime }\right)
\right) ^{2}}{2\left( t-t^{\prime }\right) }\right) \nu ^{<}\left( t^{\prime
}\right) }{\sqrt{2\pi \left( t-t^{\prime }\right) ^{3}}}dt^{\prime },%
\end{array}%
\end{equation*}%
where%
\begin{equation*}
\begin{array}{c}
-\nu ^{<}\left( t^{\prime }\right) +\int_{0}^{t}\frac{\Theta ^{<}\left(
t,t^{\prime }\right) \Xi ^{<}\left( t,t^{\prime }\right) \nu ^{<}\left(
t^{\prime }\right) }{\sqrt{2\pi \left( t-t^{\prime }\right) }}dt^{\prime
}=f^{<}\left( t\right) ,%
\end{array}%
\end{equation*}

Finally, the solution to the two-sided problem%
\begin{equation*}
\begin{array}{c}
\frac{\partial }{\partial t}F^{><}\left( t,x\right) =\frac{1}{2}\frac{%
\partial ^{2}}{\partial x^{2}}F^{><}\left( t,x\right) ,\ \ b^{>}\left(
t\right) \leq \ x\leq b^{<}\left( t\right) , \\ 
\\ 
F^{><}\left( 0,x\right) =0,\ \ \ F^{><}\left( t,b^{<}\left( t\right) \right)
=f^{<}\left( t\right) ,\ \ \ F^{><}\left( t,b^{>}\left( t\right) \right)
=f^{>}\left( t\right) ,%
\end{array}%
\end{equation*}%
has the form%
\begin{equation*}
\begin{array}{c}
F^{><}\left( t,x\right) =\int_{0}^{t}\frac{\left( x-b^{>}\left( t^{\prime
}\right) \right) \exp \left( -\frac{\left( x-b^{>}\left( t^{\prime }\right)
\right) ^{2}}{2\left( t-t^{\prime }\right) }\right) \nu ^{>}\left( t^{\prime
}\right) }{\sqrt{2\pi \left( t-t^{\prime }\right) ^{3}}}dt^{\prime
}+\int_{0}^{t}\frac{\left( x-b^{<}\left( t^{\prime }\right) \right) \exp
\left( -\frac{\left( x-b^{<}\left( t^{\prime }\right) \right) ^{2}}{2\left(
t-t^{\prime }\right) }\right) \nu ^{<}\left( t^{\prime }\right) }{\sqrt{2\pi
\left( t-t^{\prime }\right) ^{3}}}dt^{\prime },%
\end{array}%
\end{equation*}%
\begin{equation*}
\begin{array}{c}
\nu ^{>}\left( t^{\prime }\right) +\int_{0}^{t}\frac{\Theta ^{>>}\left(
t,t^{\prime }\right) \Xi ^{>>}\left( t,t^{\prime }\right) \nu ^{>}\left(
t^{\prime }\right) }{\sqrt{2\pi \left( t-t^{\prime }\right) }}dt^{\prime
}+\int_{0}^{t}\frac{\Theta ^{><}\left( t,t^{\prime }\right) \Xi ^{><}\left(
t,t^{\prime }\right) \nu ^{<}\left( t^{\prime }\right) }{\sqrt{2\pi \left(
t-t^{\prime }\right) }}dt^{\prime }=f^{>}\left( t\right) , \\ 
\\ 
-\nu ^{<}\left( t^{\prime }\right) +\int_{0}^{t}\frac{\Theta ^{<>}\left(
t,t^{\prime }\right) \Xi ^{<>}\left( t,t^{\prime }\right) \nu ^{>}\left(
t^{\prime }\right) }{\sqrt{2\pi \left( t-t^{\prime }\right) }}dt^{\prime
}+\int_{0}^{t}\frac{\Theta ^{<<}\left( t,t^{\prime }\right) \Xi ^{<<}\left(
t,t^{\prime }\right) \nu ^{<}\left( t^{\prime }\right) }{\sqrt{2\pi \left(
t-t^{\prime }\right) }}dt^{\prime }=f^{<}\left( t\right) ,%
\end{array}%
\end{equation*}%
where%
\begin{equation*}
\begin{array}{c}
\Theta ^{>>}\left( t,t^{\prime }\right) =\frac{b^{>}\left( t\right)
-b^{>}\left( t^{\prime }\right) }{\left( t-t^{\prime }\right) },\ \ \ \Theta
^{><}\left( t,t^{\prime }\right) =\frac{b^{>}\left( t\right) -b^{<}\left(
t^{\prime }\right) }{\left( t-t^{\prime }\right) },\ \ \ \text{etc.}%
\end{array}%
\end{equation*}

\subsection{Extensions\label{Extensions}}

While Eqs (\ref{Eq2}), (\ref{Eq3}) provide an elegant solution to problem (%
\ref{Eq1}), in many instances we are interested in the behavior of this
solution on the boundary itself. For instance, in numerous problems of
mathematical finance, some of which are described below, what we need to
know is the function%
\begin{equation*}
\begin{array}{c}
g^{>}(t)=\frac{1}{2}\frac{\partial }{\partial x}F^{>}(t,b^{>}(t)),%
\end{array}%
\end{equation*}%
which represent the outflow of probability from the computational domain.
This function can be calculated in two ways.

On the one hand, we can integrate the heat equation and get%
\begin{equation*}
\begin{array}{c}
\frac{d}{dt}\dint\limits_{b^{>}\left( t\right) }^{\infty }F^{>}\left(
t,x\right) dx=\dint\limits_{b^{>}\left( t\right) }^{\infty }\frac{\partial }{%
\partial t}F^{>}\left( t,x\right) dx=\frac{1}{2}\dint\limits_{b^{>}\left(
t\right) }^{\infty }\frac{\partial ^{2}}{\partial x^{2}}F^{>}\left(
t,x\right) dx=-\frac{1}{2}\frac{\partial }{\partial x}%
F^{>}(t,b^{>}(t))=-g^{>}\left( t\right) .%
\end{array}%
\end{equation*}%
Eq. (\ref{Eq2}) yields%
\begin{equation*}
\begin{array}{c}
\dint\limits_{b^{>}\left( t\right) }^{\infty }F^{>}\left( t,x\right)
dx=\int_{0}^{t}\frac{\Xi ^{>}\left( t,t^{\prime }\right) \nu ^{>}\left(
t^{\prime }\right) }{\sqrt{2\pi \left( t-t^{\prime }\right) }}dt^{\prime },%
\end{array}%
\end{equation*}%
so that%
\begin{equation}
\begin{array}{c}
g^{>}\left( t\right) =-\frac{d}{dt}\int_{0}^{t}\frac{\Xi ^{>}\left(
t,t^{\prime }\right) \nu ^{>}\left( t^{\prime }\right) }{\sqrt{2\pi \left(
t-t^{\prime }\right) }}dt^{\prime }.%
\end{array}
\label{Eq16}
\end{equation}

On the other hand, a useful formula derived by the present author and his
collaborators, see \cite{Lipton2018b, Lipton2020a, Lipton2020b, Lipton2020d}%
, gives an alternative expression for $g^{>}\left( t\right) $:%
\begin{equation}
\begin{array}{c}
g^{>}\left( t\right) =-\left( \frac{1}{\sqrt{2\pi t}}+\frac{db^{>}\left(
t\right) }{dt}\right) \nu ^{>}\left( t\right) -\frac{1}{2}\int_{0}^{t}\frac{%
\left( \ \Phi ^{>}\left( t,t^{\prime }\right) +\ \Theta ^{>2}\left(
t,t^{\prime }\right) \Xi ^{>}\left( t,t^{\prime }\right) \nu ^{>}\left(
t^{\prime }\right) \right) }{\sqrt{2\pi \left( t-t^{\prime }\right) }}%
dt^{\prime },%
\end{array}
\label{Eq17}
\end{equation}%
where%
\begin{equation*}
\begin{array}{c}
\ \ \Phi ^{>}\left( t,t^{\prime }\right) =\frac{\left( \nu ^{>}\left(
t\right) -\Xi ^{>}\left( t,t^{\prime }\right) \nu ^{>}\left( t^{\prime
}\right) \right) }{\left( t-t^{\prime }\right) },\ \ \ \ \ \Phi ^{>}\left(
t,t\right) =\frac{d\nu ^{>}\left( t\right) }{dt}+\frac{1}{2}\left( \frac{%
db^{>}\left( t\right) }{dt}\right) ^{2}\nu ^{>}\left( t\right) .%
\end{array}%
\end{equation*}

On the surface, Eqs (\ref{Eq16}), (\ref{Eq17}) look very different. However,
a useful Lemma proven in \cite{Lipton2020d}, allows one to connect the two.

\textbf{Lemma} \textit{Let }$\Psi \left( t,t^{\prime }\right) $\textit{\ be
a differentiable function, such that }$\Psi \left( t,t\right) =1$\textit{.
Then}%
\begin{equation*}
\begin{array}{c}
\frac{d}{dt}\int_{0}^{t}\frac{\Psi \left( t,t^{\prime }\right) \nu \left(
t^{\prime }\right) }{\sqrt{2\pi \left( t-t^{\prime }\right) }}dt^{\prime }=%
\frac{\nu \left( t\right) }{\sqrt{2\pi t}}+\frac{1}{2}\int_{0}^{t}\frac{\nu
\left( t\right) -\left( \Psi \left( t,t^{\prime }\right) -2\left(
t-t^{\prime }\right) \Psi _{t}\left( t,t^{\prime }\right) \right) \nu \left(
t^{\prime }\right) }{\sqrt{2\pi \left( t-t^{\prime }\right) ^{3}}}dt^{\prime
},%
\end{array}%
\end{equation*}%
\textit{Alternatively,}%
\begin{equation*}
\begin{array}{c}
\frac{d}{dt}\int_{0}^{t}\frac{\Psi \left( t,t^{\prime }\right) \nu \left(
t^{\prime }\right) }{\sqrt{2\pi \left( t-t^{\prime }\right) }}dt^{\prime
}=\int_{0}^{t}\frac{\frac{\partial }{\partial t^{\prime }}\left( \left( \Psi
\left( t,t^{\prime }\right) -2\left( t-t^{\prime }\right) \Psi _{t}\left(
t,t^{\prime }\right) \right) \nu \left( t^{\prime }\right) \right) }{\sqrt{%
2\pi \left( t-t^{\prime }\right) }}dt^{\prime }.%
\end{array}%
\end{equation*}

We emphasize that Eq. (\ref{Eq17}) is easier to use than Eq. (\ref{Eq16}) in
most situations because it does not involve differentiation. However, if the
cumulative outflow $G^{>}\left( t\right) =\int_{0}^{t}g^{>}\left( t^{\prime
}\right) dt^{\prime }$ is of interest, the latter equation can be more
efficient, since it can be rewritten as follows: 
\begin{equation*}
\begin{array}{c}
G^{>}\left( t\right) =-\int_{0}^{t}\frac{\Xi ^{>}\left( t,t^{\prime }\right)
\nu ^{>}\left( t^{\prime }\right) }{\sqrt{2\pi \left( t-t^{\prime }\right) }}%
dt^{\prime }.%
\end{array}%
\end{equation*}

We can calculate $g^{<}(t)$ and $g^{><}(t)$ by the same token. It is
important to understand that both Eqs (\ref{Eq17}) and (\ref{Eq16}) can be
used in the one-sided case, however, in the case when two boundaries are
present, we can \emph{only} use Eq. (\ref{Eq17}) because this equation
allows calculating $g^{>}$ and $g^{<}$ individually while Eq. (\ref{Eq16})
calculates the difference $g^{>}-g^{<}$.

\subsection{Generalizations}

If the MHP were applicable only to standard Wiener process, it would be
advantageous, if somewhat narrow in scope. Fortunately, it can be applied to
a general diffusion satisfying the so-called Cherkasov's condition, which
guarantees that it can be transformed into the standard Wiener process. Such
diffusions are studied in \cite{Cherkasov1957}, \cite{Ricciardi1976}, and 
\cite{Bluman1980}. The applications of Cherkasov's condition in financial
mathematics are discussed in \cite{Lipton2001}, Section 4.2, and \cite%
{Lipton2018a}, Chapter 9.

Consider a diffusion governed by%
\begin{equation*}
\begin{array}{c}
d\tilde{x}_{\tilde{t}}=\delta \left( \tilde{t},\tilde{x}_{\tilde{t}}\right)
\,d\tilde{t}+\sigma \left( \tilde{t},\tilde{x}_{\tilde{t}}\right) \,dW_{%
\tilde{t}},\ \ \ \ \ \tilde{x}_{0}=\tilde{z},%
\end{array}%
\end{equation*}%
We wish to calculate boundary-related quantities, such as the distribution
of the hitting time of a given time-dependent barrier $b\left( \tilde{t}%
\right) $: 
\begin{equation*}
\begin{array}{c}
\tilde{s}=\inf \left\{ \tilde{t}:\tilde{x}_{\tilde{t}}=\tilde{b}(\tilde{t}%
)\right\} ,\ \ \ \ \ \tilde{z}\neq \tilde{b}(0).%
\end{array}%
\end{equation*}%
To this end, we introduce%
\begin{equation*}
\begin{array}{c}
\beta \left( \tilde{t},\tilde{x}\right) =\sigma \left( \tilde{t},\tilde{x}%
\right) \int^{\tilde{x}}\frac{1}{\sigma \left( \tilde{t},y\right) }dy, \\ 
\\ 
\gamma \left( \tilde{t},\tilde{x}\right) =2\delta \left( \tilde{t},\tilde{x}%
\right) -\sigma \left( \tilde{t},\tilde{x}\right) \sigma _{\tilde{x}}\left( 
\tilde{t},\tilde{x}\right) -2\sigma \left( \tilde{t},\tilde{x}\right) \int^{%
\tilde{x}}\frac{\sigma _{\tilde{t}}\left( \tilde{t},y\right) }{\sigma
^{2}\left( \tilde{t},y\right) }dy,%
\end{array}%
\end{equation*}%
where the lower limit of integration is chosen as convenient. Define%
\begin{equation*}
\begin{array}{c}
P\left( \tilde{t},\tilde{x}\right) =\left\vert 
\begin{array}{cc}
\beta \left( \tilde{t},\tilde{x}\right) & \gamma \left( \tilde{t},\tilde{x}%
\right) \\ 
\beta _{\tilde{x}}\left( \tilde{t},\tilde{x}\right) & \gamma _{\tilde{x}%
}\left( \tilde{t},\tilde{x}\right)%
\end{array}%
\right\vert , \\ 
\\ 
Q\left( \tilde{t},\tilde{x}\right) =\left\vert 
\begin{array}{cc}
\sigma \left( \tilde{t},\tilde{x}\right) & \gamma \left( \tilde{t},\tilde{x}%
\right) \\ 
\sigma _{\tilde{x}}\left( \tilde{t},\tilde{x}\right) & \gamma _{\tilde{x}%
}\left( \tilde{t},\tilde{x}\right)%
\end{array}%
\right\vert , \\ 
\\ 
R\left( \tilde{t},\tilde{x}\right) =\left\vert 
\begin{array}{ccc}
\sigma \left( \tilde{t},\tilde{x}\right) & \beta \left( \tilde{t},\tilde{x}%
\right) & \gamma \left( \tilde{t},\tilde{x}\right) \\ 
\sigma _{\tilde{x}}\left( \tilde{t},\tilde{x}\right) & \beta _{\tilde{x}%
}\left( \tilde{t},\tilde{x}\right) & \gamma _{\tilde{x}}\left( \tilde{t},%
\tilde{x}\right) \\ 
\sigma _{\tilde{x}\tilde{x}}\left( \tilde{t},\tilde{x}\right) & \beta _{%
\tilde{x}\tilde{x}}\left( \tilde{t},\tilde{x}\right) & \gamma _{\tilde{x}%
\tilde{x}}\left( \tilde{t},\tilde{x}\right)%
\end{array}%
\right\vert ,%
\end{array}%
\end{equation*}%
and assume that Cherkasov's condition is satisfied, so that%
\begin{equation*}
\begin{array}{c}
R\left( \tilde{t},\tilde{x}\right) \equiv 0.%
\end{array}%
\end{equation*}%
Then we can transform $\tilde{x}$ into the standard Wiener process via the
following mapping%
\begin{equation*}
\begin{array}{c}
t=t(\tilde{t},\tilde{x})=\int_{0}^{\tilde{t}}\Phi ^{2}(u,\tilde{x})du, \\ 
\\ 
x=x(\tilde{t},\tilde{x})=\Phi (\tilde{t},\tilde{x})\frac{\beta \left( \tilde{%
t},\tilde{x}\right) }{\sigma \left( \tilde{t},\tilde{x}\right) }+\frac{1}{2}%
\int_{0}^{\tilde{t}}\Phi (u,\tilde{x})\frac{P\left( u,\tilde{x}\right) }{%
\sigma \left( u,\tilde{x}\right) }du,%
\end{array}%
\end{equation*}%
where 
\begin{equation*}
\begin{array}{c}
\Phi (\tilde{t},\tilde{x})=\exp \left[ -\frac{1}{2}\int_{0}^{t}\frac{Q\left(
u,\tilde{x}\right) }{\sigma \left( u,\tilde{x}\right) }du\right] .%
\end{array}%
\end{equation*}%
In particular, the initial condition becomes 
\begin{equation*}
\begin{array}{c}
z=\frac{\beta \left( 0,\tilde{z}\right) }{\sigma \left( 0,\tilde{z}\right) }.%
\end{array}%
\end{equation*}%
The corresponding transition probability density transforms as follows 
\begin{equation*}
\begin{array}{c}
\tilde{p}(\tilde{t},\tilde{x};\tilde{z})=\left\vert \frac{\partial x(\tilde{t%
},\tilde{x})}{\partial \tilde{x}}\right\vert p(t,x;z).%
\end{array}%
\end{equation*}%
Moreover, the boundary transforms to 
\begin{equation*}
\begin{array}{c}
\tilde{b}(\tilde{t})\rightarrow b(t)=\Phi (\tilde{t},\tilde{b}(\tilde{t}))%
\frac{\beta \left( \tilde{t},\tilde{b}(\tilde{t})\right) }{\sigma \left( 
\tilde{t},\tilde{b}(\tilde{t})\right) }+\frac{1}{2}\int^{\tilde{t}}\Phi (u,%
\tilde{b}(\tilde{t}))\frac{P\left( u,\tilde{b}(\tilde{t})\right) }{\sigma
\left( u,\tilde{b}(\tilde{t})\right) }du.%
\end{array}%
\end{equation*}

Since the MHP\ is specifically designed for dealing with curvilinear
boundaries, we get a solvable problem. A powerful application of the above
approach is demonstrated in Section \ref{OU}, where the hitting time
probability distribution for an Ornstein-Uhlenbeck process is studied.

\subsection{Numerics}

There are numerous well-known approaches to solving Volterra equations; see, 
\cite{Linz1985}, among many others. We choose the most straightforward
approach and show how to solve the following archetypal Volterra equation
with weak singularity numerically: 
\begin{equation*}
\begin{array}{c}
\nu (t)+\int_{0}^{t}\frac{K(t,t^{\prime })}{\sqrt{t-t^{\prime }}}\nu
(t^{\prime })\,dt^{\prime }=f(t),%
\end{array}%
\end{equation*}%
where $K(t,t^{\prime })$ is a non-singular kernel. We write

\begin{equation}
\begin{array}{c}
\int_{0}^{t}\frac{K(t,t^{\prime })\nu \left( t^{\prime }\right) }{\sqrt{%
t-t^{\prime }}}dt^{\prime }=-2\int_{0}^{t}K(t,t^{\prime })\nu \left(
t^{\prime }\right) \,d\sqrt{t-t^{\prime }}.%
\end{array}
\label{Eq22}
\end{equation}%
We wish to map this equation to a grid $0=t_{0}<t_{1}<\ldots <t_{N}=T$. To
this end, we introduce the following notation: 
\begin{equation*}
\begin{array}{c}
f_{k}=f(t_{k}),\ \ \ \nu _{k}=\nu \left( t_{k}\right) ,\ \ \
K_{k,l}=K(t_{k},t_{l}),\ \ \ \Delta _{k,l}=t_{k}-t_{l}.%
\end{array}%
\end{equation*}%
Then, the right hand side of Eq. (\ref{Eq22}) can be approximated by the
trapezoidal rule as

\begin{equation}
\begin{array}{c}
f_{k}=\nu _{k}+\sum_{l=1}^{k}\left( K_{k,l}\nu _{l}+K_{k,l-1}\nu
_{l-1}\right) \Pi _{k,l}=0,%
\end{array}
\label{Eq24}
\end{equation}%
where%
\begin{equation*}
\begin{array}{c}
\Pi _{k,l}=\frac{\Delta _{l,l-1}}{\left( \sqrt{\Delta _{k,l-1}}+\sqrt{\Delta
_{k,l}}\right) },%
\end{array}%
\end{equation*}%
so that

\begin{equation}
\begin{array}{c}
\nu _{k}=\frac{\left( f_{k}-K_{k,k-1}\nu _{k-1}-\sum_{l=1}^{k-1}\left(
K_{k,l}\nu _{l}+K_{k,l-1}\nu _{l-1}\right) \Pi _{k,l}\right) }{\left(
1+K_{k,k}\sqrt{\Delta _{k,k-1}}\right) }.%
\end{array}
\label{Eq25}
\end{equation}%
Thus, $\nu _{k}$ can be found by induction starting with $\nu _{0}=f_{0}$.

Eq. (\ref{Eq25}) is the blueprint for all the subsequent numerical
calculations.

\section{The structural default model\label{Strudefa}}

\subsection{Preliminaries}

The original, and straightforward, structural default model was introduced
by Merton, \cite{Merton1974}, who assumed that default could happen only at
debt maturity. His model was extended by Black and Cox, \cite{Black1976},
who considered the default, which can happen at any time by introducing flat
default boundary representing debt covenants. Numerous authors expanded
their model including \cite{Hyer1998, Hull2001, Avellaneda2001}, who
considered curvilinear boundary whose shape can be calibrated to the market
default probability. One of the major unsolved issues with the above model
was articulated by Hyer \textit{et al.}, \cite{Hyer1998}, who pointed out
that, unless the shape of the default boundary is very carefully chosen, the
probability of short-term default is too low. This issue was addressed by
several authors, including \cite{Finkelstein2001, Hilberink2002, Lipton2002}%
, who proposed to introduce jump and or uncertainty to increase this
probability. We show below that it is possible to calibrate the default
boundary in such a way that constant default intensity can be matched. We
emphasize that the direct problem - calculating the default probability
given the boundary - is linear (albeit relatively involved), while the
inverse problem - finding the boundary given the default probability - is
nonlinear (and hence even more involved). Additional details are given in 
\cite{Lipton2020b}.

\subsection{Formulation}

We wish to find the boundary for a structural default model corresponding to
a constant default intensity $\eta $. We denote the corresponding default
probability by 
\begin{equation*}
\begin{array}{c}
\pi \left( t\right) =1-e^{-\eta t}.%
\end{array}%
\end{equation*}%
The introduce time $\tau $, such that default is impossible for $t<\tau $.
Thus the default boundary starts at $t=\tau $. The idea is to calculate the
corresponding boundary $b\left( t;\tau ,\eta \right) $, provided it exists,
and then let $\tau \rightarrow 0$.

It is clear that at time $t=\tau -0$, the transition probability is 
\begin{equation*}
\begin{array}{c}
p(\tau ,x)=H\left( \tau ,x\right) ,%
\end{array}%
\end{equation*}%
where $H$ is the heat kernel:%
\begin{equation*}
\begin{array}{c}
H\left( \tau ,x\right) =\frac{e^{-\frac{x^{2}}{2\tau }}}{\sqrt{2\pi \tau }}.%
\end{array}%
\end{equation*}%
At time $t=\tau $, the first possibility of default occurs. For $t>\tau $
the transition probability satisfies the following Fokker--Planck problem 
\begin{equation*}
\begin{array}{c}
\frac{\partial }{\partial t}p(t,x)=\frac{1}{2}\frac{\partial ^{2}}{\partial
x^{2}}p(t,x),\ \ \ \ \ b\left( t\right) \leq x<\infty , \\ 
\\ 
p\left( \tau ,x\right) =H\left( \tau ,x\right) ,\ \ \ p\left( t,b\left(
t\right) \right) =0,\ \ \ p\left( t,x\rightarrow \infty \right) \rightarrow
0.%
\end{array}%
\end{equation*}%
The default probability density $g(t)$ is given by%
\begin{equation*}
\begin{array}{c}
g(t)=\frac{1}{2}\frac{\partial }{\partial x}p(t,b(t)).%
\end{array}%
\end{equation*}%
Alternatively, 
\begin{equation*}
\begin{array}{c}
\pi (t)=1-\int_{b(t)}^{\infty }p(t,x)\,dx,\ \ \ g(t)=\frac{d\pi \left(
t\right) }{dt}.%
\end{array}%
\end{equation*}

\subsection{Governing system of integral equations}

We split $p$ as follows%
\begin{equation*}
p(t,x)=q(t,x)+r(t,x),
\end{equation*}%
where 
\begin{equation}
\begin{array}{c}
\frac{\partial }{\partial t}q(t,x)=\frac{1}{2}\frac{\partial ^{2}}{\partial
x^{2}}q(t,x),\ \ \ \ \ -\infty <x<\infty , \\ 
\\ 
q\left( \tau ,x\right) =H\left( \tau ,x\right) \Theta \left( x-b\left( \tau
\right) \right) ,\ \ \ q\left( t,x\rightarrow -\infty \right) \rightarrow
0,\ \ \ q\left( t,x\rightarrow \infty \right) \rightarrow 0,%
\end{array}
\label{Eq31}
\end{equation}%
\begin{equation*}
\begin{array}{c}
\frac{\partial }{\partial t}r(t,x)=\frac{1}{2}\frac{\partial ^{2}}{\partial
x^{2}}r(t,x),\ \ \ \ \ b\left( t\right) \leq x<\infty , \\ 
\\ 
r\left( \tau ,x\right) =0,\ \ \ r\left( t,b\left( t\right) \right) =-q\left(
t,b\left( t\right) \right) ,\ \ \ r\left( t,x\rightarrow \infty \right)
\rightarrow 0.%
\end{array}%
\end{equation*}%
and $\Theta \left( x\right) $ is the Heaviside function. Solving Eq. (\ref%
{Eq31}) as a convolution of heat kernel with the initial condition, we get 
\begin{equation*}
\begin{array}{c}
q(t,x)=\frac{e^{-\frac{x^{2}}{2t}}}{\sqrt{2\pi t}}N\left( \frac{\frac{ux}{t}%
-b\left( \tau \right) }{\sqrt{u}}\right) ,%
\end{array}%
\end{equation*}%
where $u=(t-\tau )\tau /t$, see \cite{Lipton2020b}. Thus%
\begin{equation*}
\begin{array}{c}
g\left( t\right) =\frac{1}{2}\frac{\partial }{\partial x}r\left( t,b\left(
t\right) \right) -\frac{H\left( t,b\left( t\right) \right) }{2t}\left(
b\left( t\right) N\left( \frac{\frac{ub\left( t\right) }{t}-b\left( \tau
\right) }{\sqrt{u}}\right) -uH\left( u,\frac{ub\left( t\right) }{t}-b\left(
\tau \right) \right) \right) .%
\end{array}%
\end{equation*}%
\qquad

Accordingly, in view the discussion in Section \ref{Extensions}, we need to
solve the following system of integral equations:%
\begin{equation}
\begin{array}{c}
\nu \left( t\right) +\int_{\tau }^{t}\frac{\Theta \left( t,t^{\prime
}\right) \Xi \left( t,t^{\prime }\right) \nu \left( t^{\prime }\right) }{%
\sqrt{2\pi \left( t-t^{\prime }\right) }}dt^{\prime }+H\left( t,b\left(
t\right) \right) N\left( \frac{ub\left( t\right) -tb\left( \tau \right) }{t%
\sqrt{u}}\right) =0, \\ 
\\ 
\eta e^{-\eta t}+\left( \frac{1}{\sqrt{2\pi t}}+\frac{db\left( t\right) }{dt}%
\right) \nu \left( t\right) +\frac{1}{2}\int_{\tau }^{t}\frac{\Phi \left(
t,t^{\prime }\right) +\Theta ^{2}\left( t,t^{\prime }\right) \Xi \left(
t,t^{\prime }\right) \nu \left( t^{\prime }\right) }{\sqrt{2\pi \left(
t-t^{\prime }\right) }}dt^{\prime } \\ 
\\ 
+\frac{H\left( t,b\left( t\right) \right) }{2t}\left( b\left( t\right)
N\left( \frac{\frac{ub\left( t\right) }{t}-b\left( \tau \right) }{\sqrt{u}}%
\right) -uH\left( u,\frac{ub\left( t\right) }{t}-b\left( \tau \right)
\right) \right) =0.%
\end{array}
\label{Eq35}
\end{equation}

Alternatively, we can rewrite Eqs (\ref{Eq35}) in integrated form%
\begin{equation}
\begin{array}{c}
\nu \left( t\right) +\int_{\tau }^{t}\frac{\Theta \left( t,t^{\prime
}\right) \Xi \left( t,t^{\prime }\right) \nu \left( t^{\prime }\right) }{%
\sqrt{2\pi \left( t-t^{\prime }\right) }}dt^{\prime }+H\left( t,b\left(
t\right) \right) N\left( \frac{ub\left( t\right) -tb\left( \tau \right) }{t%
\sqrt{u}}\right) =0, \\ 
\\ 
1-e^{-\eta t}+\int_{\tau }^{t}\frac{\Xi \left( t,t^{\prime }\right) \nu
\left( t^{\prime }\right) }{\sqrt{2\pi \left( t-t^{\prime }\right) }}%
dt^{\prime }-N\left( \frac{b\left( t\right) }{\sqrt{t}}\right) -N\left( 
\frac{\sqrt{t}b\left( \tau \right) }{\sqrt{u\left( u+t\right) }}\right)
+BVN\left( \frac{\sqrt{t}b\left( \tau \right) }{\sqrt{u\left( u+t\right) }},%
\frac{b\left( t\right) }{\sqrt{t}};\sqrt{\frac{u}{u+t}}\right) =0,%
\end{array}
\label{Eq36}
\end{equation}%
where $BVN\left( .,.;.\right) $ is the bivariate normal distribution.

We postpone the discussion of the corresponding numerics until the next
Section, where a more general case is considered.

\subsection{The choice of $b_{\protect\tau }$}

Recall that the default probability has the form 
\begin{equation*}
\begin{array}{c}
\pi (t)=1-e^{-\eta t}.%
\end{array}%
\end{equation*}%
The barrier has to start at $\tau =\hat{\tau}$, $\hat{\tau}\rightarrow 0$,
and there should be no barrier before that. We wish to find $b\left( \hat{%
\tau}\right) $ such that%
\begin{equation*}
\begin{array}{c}
\pi \left( \hat{\tau}\right) =1-\int_{b\left( \hat{\tau}\right) }^{\infty }%
\frac{\exp \left( -\frac{x^{2}}{2\hat{\tau}}\right) }{\sqrt{2\pi \hat{\tau}}}%
dx=1-N\left( -\frac{b\left( \hat{\tau}\right) }{\sqrt{\hat{\tau}}}\right)
=1-e^{-\eta \hat{\tau}}.%
\end{array}%
\end{equation*}%
Thus,%
\begin{equation*}
\begin{array}{c}
N\left( -\frac{b\left( \hat{\tau}\right) }{\sqrt{\hat{\tau}}}\right)
=e^{-\eta \hat{\tau}},%
\end{array}%
\end{equation*}%
and%
\begin{equation*}
\begin{array}{c}
b\left( \hat{\tau}\right) =-\sqrt{\hat{\tau}}N^{-1}\left( e^{-\eta \hat{\tau}%
}\right) .%
\end{array}%
\end{equation*}%
Now,%
\begin{equation*}
\begin{array}{c}
N^{-1}\left( y\right) \underset{y\rightarrow 1}{\sim }\sqrt{2f\left( \eta
\right) },%
\end{array}%
\end{equation*}%
where%
\begin{equation*}
\begin{array}{c}
\eta =-\ln \left( 2\sqrt{\pi }\left( 1-y\right) \right) ,\ \ \ f\left( \eta
\right) =\eta -\frac{\ln \eta }{2}+\frac{\ln \eta -2}{4\eta }+\frac{\left(
\ln \eta \right) ^{2}-6\ln \eta +14}{16\eta ^{2}},%
\end{array}%
\end{equation*}%
so that%
\begin{equation*}
\begin{array}{c}
b\left( \hat{\tau}\right) =-\sqrt{2\hat{\tau}f\left( -\ln \left( 2\sqrt{\pi }%
\left( 1-e^{-\eta \hat{\tau}}\right) \right) \right) }\approx -\sqrt{2\hat{%
\tau}\ln \left( \frac{1}{2\sqrt{\pi }\eta \hat{\tau}}\right) }.%
\end{array}%
\end{equation*}

\subsection{Default boundaries}

\begin{figure}[tbp]
\begin{center}
\subfloat[]{\includegraphics[width=0.5\textwidth]
{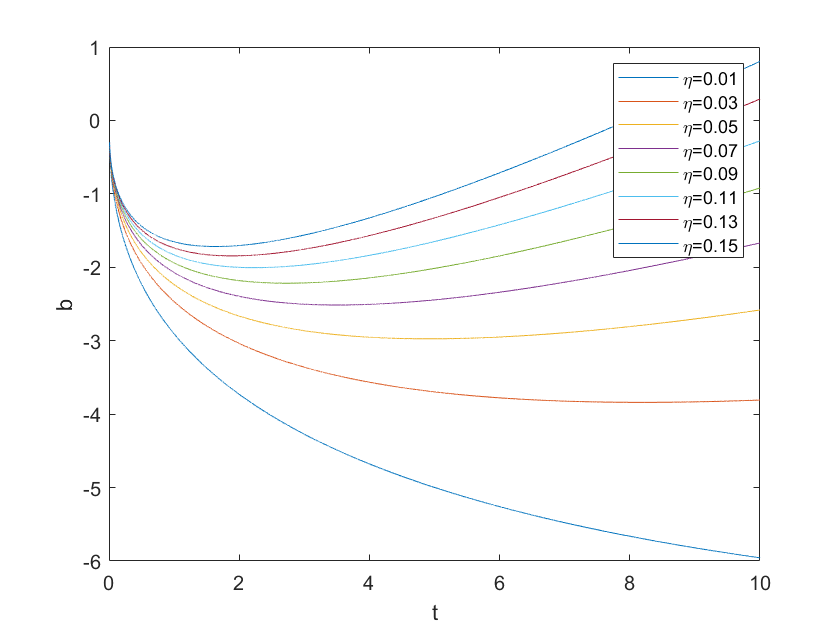}}
\\[0pt]
\subfloat[]{\includegraphics[width=0.5\textwidth]
{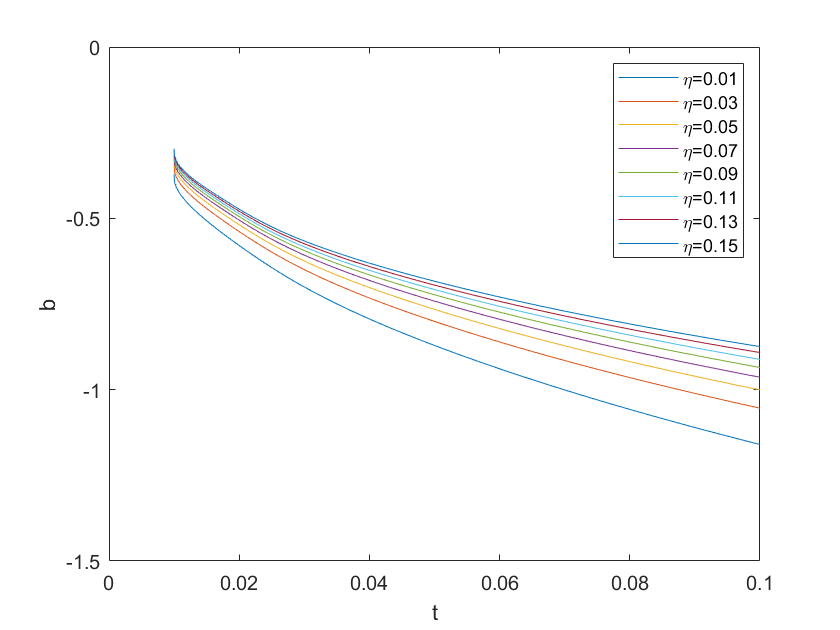}}%
\\[0pt]
\end{center}
\par
\vspace{-10pt}
\caption{In Figures (a)-(b), we show the default boundaries for several
representative values of the default intensity $\protect\eta $. We choose $%
\protect\tau =0.01$. In Figure (a), we choose $0.01<t<10.0$ to capture their
overall behavior; In Figure (b), we choose $0.01<t<0.1$ so that small
features can be shown.}
\label{Fig1}
\end{figure}

\begin{figure}[tbp]
\begin{center}
{%
\includegraphics[width=1.0\textwidth]
{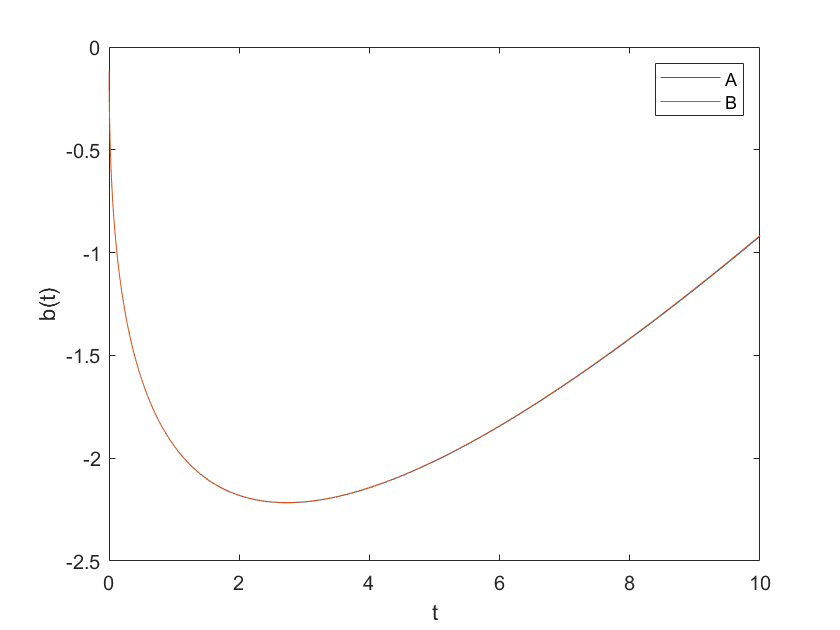}%
}
\end{center}
\par
\vspace{-10pt}
\caption{In this figure, we choose the default intensity $\protect\eta =0.09$
and show that boundaries calculated by solving Eqs (\protect\ref{Eq35}) and
Eqs (\protect\ref{Eq36}) coincide modulo numerical errors.}
\label{Fig2}
\end{figure}

Default boundaries calibrated to several representative values of $\eta $
are shown in Figure \ref{Fig1}.%
\begin{equation*}
\text{Figure \ref{Fig1} near here.}
\end{equation*}%
We show that solutions of Eqs (\ref{Eq35}) and Eqs (\ref{Eq36}) coincide
modulo numerical error in Figure \ref{Fig2}.%
\begin{equation*}
\text{Figure \ref{Fig2} near here.}
\end{equation*}

\subsection{Main conjecture}

\textbf{Conjecture} \textit{For a given time interval }$I^{\left( T\right) }=%
\left[ 0,T\right] $\textit{, there exists a parameter interval }$I^{\left(
\eta \right) }\left( T\right) =\left[ 0,\eta ^{\ast }\left( T\right) \right] 
$\textit{, such that for any }$\eta \in I^{\left( \eta \right) }\left(
T\right) $\textit{, the default boundary }$b\left( t\right) $\textit{\ can
be calibrated to the default intensity }$\eta $\textit{. We can construct
the corresponding boundary as follows:}%
\begin{equation*}
\begin{array}{c}
b\left( t;\eta \right) =\lim_{\tau \rightarrow 0}b\left( t;\tau ,\eta
\right) ,\ \ \ 0<t\leq T,%
\end{array}%
\end{equation*}%
\textit{where }$b\left( t;\tau ,\eta \right) $\textit{\ is found by solving
either Eqs (\ref{Eq35}) or Eqs (\ref{Eq36}).}

\begin{figure}[tbp]
\begin{center}
\subfloat[]{\includegraphics[width=0.5\textwidth]
{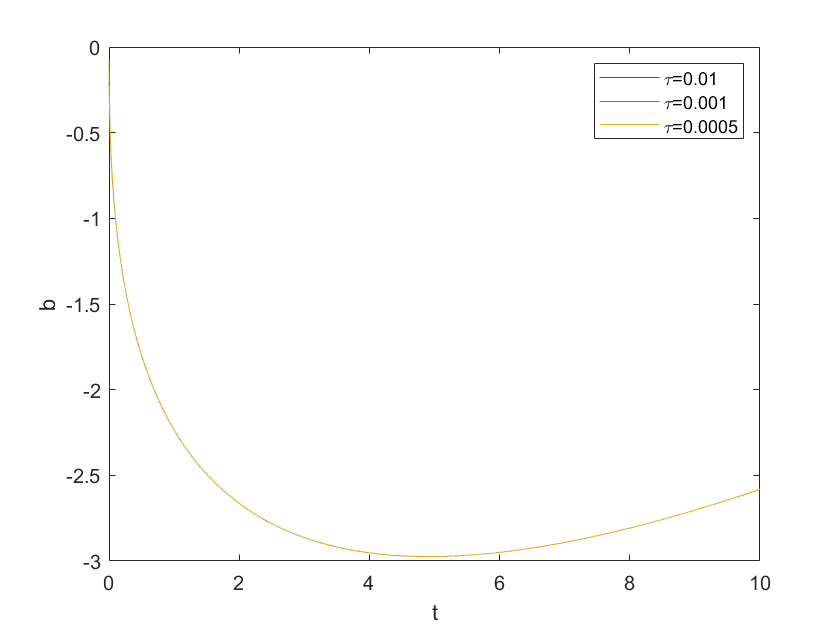}}
\\[0pt]
\subfloat[]{\includegraphics[width=0.5\textwidth]
{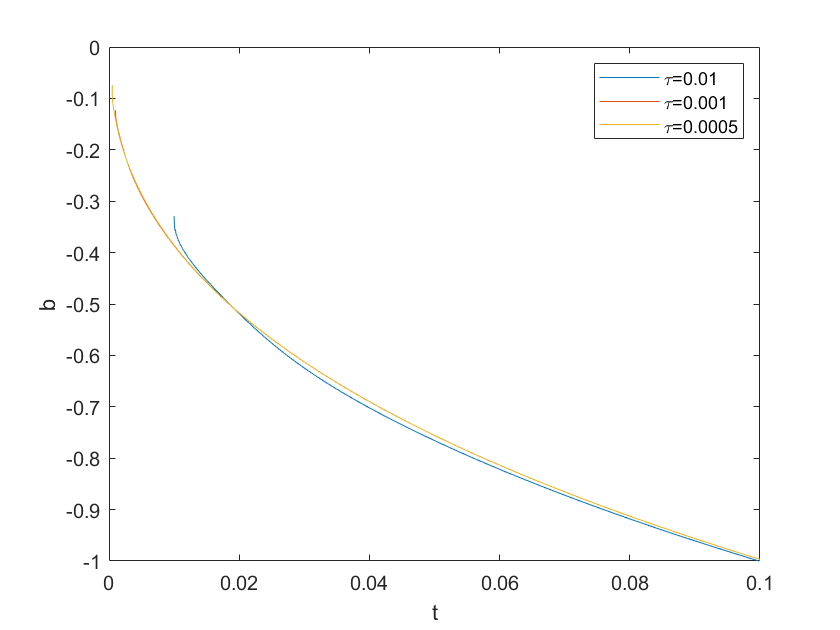}}%
\\[0pt]
\subfloat[]{\includegraphics[width=0.5\textwidth]
{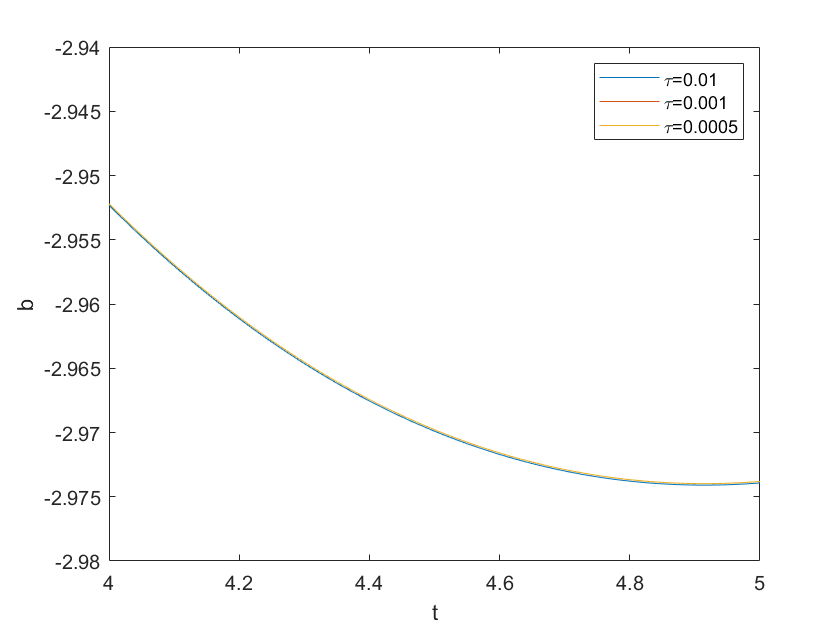}}
\end{center}
\par
\vspace{-10pt}
\caption{In this figure, we choose the default intensity $\protect\eta =0.05$
and illustrate our main conjecture numerically by constructing three
boundaries corresponding to $\protect\tau =0.01$, $0.001$, and $0.0005$,
respectively. It is clear that after a short initial period, these
boundaries begin overlapping.}
\label{Fig3}
\end{figure}

We illustrate our conjecture in Figure \ref{Fig3}.%
\begin{equation*}
\text{Figure \ref{Fig3} near here.}
\end{equation*}

\section{Mean-field banking system}

\subsection{Preliminaries}

No bank is an island - they operate as a group. Tangible links between banks
manifest themselves via interbank loans; intangible links are manifold -
overall sentiment, ease of doing business, and others. Hence, to build a
meaningful structural default model for a bank, one needs to take into
account this bank's interactions with all the banks whom it lends to or
borrows from. Eisenberg and Noe developed a Merton-like model of the bank
default (default can happen only at maturity) in the seminal paper \cite%
{Eisenberg2001}. The present author extended the Eisenberg-Noe model to the
Black-Cox setting (default can happen at any time before maturity provided
that debt covenants are violated); see \cite{Lipton2016}. Lipton's work was
subsequently generalized in \cite{Itkin2015, Itkin2017}. Recently, several
authors considered the interconnected banking system in the mean-field
framework and studied a representative bank, see \cite{Hambly2018,
Ichiba2018, Kaushansky2018, Nadtochiy2017, Nadtochiy2018} among many others.
In this section, we also use the mean-field approach. Additional details are
given in \cite{Lipton2020d}.

\subsection{Interconnected banking system}

We follow \cite{Lipton2016} and assume that the dynamics of bank $i$'s total
external assets is governed by 
\begin{equation*}
\begin{array}{c}
\frac{dA_{t}^{i}}{A_{t}^{i}}=\mu _{i}\,dt+\sigma _{i}\,dW_{t}^{i},%
\end{array}%
\end{equation*}%
where $W^{i}$ are independent standard Brownian motions for $1\leq i\leq n$,
and the liabilities, both external $L_{i}$ and mutual $L_{ij}$, are constant.

Bank $i$ is assumed to default when its assets fall below a certain
threshold determined by its liabilities, namely at time $\tau _{i}=\inf
\{t:A_{t}^{i}\leq \Lambda _{t}^{i}\}$, where $\Lambda ^{i}$ is a default
boundary which we now work out. At time $t=0$, 
\begin{equation*}
\begin{array}{c}
\Lambda _{0}^{i}=R_{i}\left( L_{i}+\sum_{j\neq i}L_{ij}\right) -\sum_{j\neq
i}L_{ji},%
\end{array}%
\end{equation*}%
where $R_{i}$ is the recovery rate of bank $i$. If bank $k$ defaults at time 
$t$, the default boundary of bank $i$ jumps by $\Delta \Lambda
_{t}^{i}=(1-R_{i}R_{k})L_{ki}.$

The distance to default $Y_{t}^{i}=\log (A_{t}^{i}/\Lambda _{t}^{i})/\sigma $
has the following dynamics: 
\begin{equation*}
\begin{array}{c}
Y_{t}^{i}=Y_{0}^{i}+(\mu -\sigma ^{2}/2)t+W_{t}^{i}-\frac{1}{\sigma }\log
\left( 1+\frac{\gamma }{N}\sum_{k\neq i}(1-R^{2})\frac{1}{\Lambda _{0}}%
\mathbf{1}_{\{\tau _{k}\leq t\}}\right) ,%
\end{array}%
\end{equation*}%
or, approximately, 
\begin{equation*}
\begin{array}{c}
Y_{t}^{i}=Y_{t}^{0}+(\mu -\sigma ^{2}/2)t+W_{t}^{i}-\frac{\gamma (1-R^{2})}{%
\sigma \Lambda _{0}}L_{t}^{N},%
\end{array}%
\end{equation*}%
where 
\begin{equation*}
\begin{array}{c}
L_{t}^{N}=\frac{1}{N}\sum_{k}\mathbf{1}_{\{\tau _{k}\leq t\}}.%
\end{array}%
\end{equation*}%
In the limit for $N\rightarrow \infty $, all $Y_{t}^{i}$ have the same
dynamics:%
\begin{equation*}
\begin{array}{c}
Y_{t}=Y_{0}+W_{t}-\alpha L_{t}, \\ 
\\ 
L_{t}=\mathbb{P}(\tau \leq t),\ \ \ \tau =\inf \{t\in \lbrack 0,T]:Y_{t}\leq
0\},%
\end{array}%
\end{equation*}%
where $\alpha =\left. \gamma (1-R^{2})\right/ \sigma \Lambda _{0}$
characterizes the strength of interbank interactions. Thus, we are dealing
with a mean-field problem - the behavior of a representative bank depends on
the behavior of all other banks, and all of them have the same dynamics.
Hence, the problem in question is nonlinear.

We follow \cite{Lipton2020d} and write the increasing process $L$ as 
\begin{equation*}
\begin{array}{c}
\alpha L_{t}=-\int_{0}^{t}\mu (t^{\prime })\,dt^{\prime }=-M\left( t\right) ,%
\end{array}%
\end{equation*}%
for some negative $\mu $, so that $p$ satisfies

\begin{equation}
\begin{array}{c}
\frac{\partial }{\partial t}p\left( t,x;z\right) =-\mu (t)\,\frac{\partial }{%
\partial x}p\left( t,x;z\right) +\frac{1}{2}\frac{\partial ^{2}}{\partial
x^{2}}p\left( t,x;z\right) ,\ \ \ 0\leq x<\infty , \\ 
\\ 
p\left( 0,x;z\right) =\delta _{z}\left( x\right) ,\ \ \ p\left( t,0;z\right)
=0,\ \ \ p\left( t,x\rightarrow \infty \right) \rightarrow 0.%
\end{array}
\label{Eq52}
\end{equation}

As we already know, 
\begin{equation*}
\begin{array}{c}
g(t;z)\equiv \frac{dL_{t}}{dt}=\frac{1}{2}p_{x}(t,0;z),%
\end{array}%
\end{equation*}%
so that Eqs (\ref{Eq52}) can be written in the self-consistent form 
\begin{equation*}
\begin{array}{c}
\frac{\partial }{\partial t}p\left( t,x;z\right) =\frac{\alpha }{2}\frac{%
\partial }{\partial x}p(t,0;z)\frac{\partial }{\partial x}\,p\left(
t,x;z\right) +\frac{1}{2}\frac{\partial ^{2}}{\partial x^{2}}p\left(
t,x;z\right) ,\ \ \ 0\leq x<\infty , \\ 
\\ 
p\left( 0,x;z\right) =\delta _{z}\left( x\right) ,\ \ \ p\left( t,0;z\right)
=0,\ \ \ p\left( t,x\rightarrow \infty \right) \rightarrow 0.%
\end{array}%
\end{equation*}

The change of variables $\left( t,x\right) \rightarrow \left( t,y\right)
=\left( t,x-M\left( t\right) \right) $ yields the familiar
initial-boundary-value problem (IBVP):%
\begin{equation*}
\begin{array}{c}
\frac{\partial }{\partial t}p\left( t,y\right) =\frac{1}{2}p_{yy}\left(
t,y\right) ,\ \ \ \ 0\leq y<\infty , \\ 
\\ 
p\left( 0,y\right) =\delta _{z}\left( y\right) ,\ \ \ p\left( t,-M\left(
t\right) \right) =0,\ \ \ p\left( t,y\rightarrow \infty \right) \rightarrow
0.%
\end{array}%
\end{equation*}%
As before, we split $p$ in two parts%
\begin{equation*}
\begin{array}{c}
p\left( t,y\right) =H\left( t,y\right) +r\left( t,y\right) ,%
\end{array}%
\end{equation*}%
where $H\left( t,y\right) $ is the standard heat kernel, while $r$ is the
solution of the following problem:%
\begin{equation*}
\begin{array}{c}
\frac{\partial }{\partial t}r\left( t,y\right) =\frac{1}{2}\frac{\partial
^{2}}{\partial y^{2}}r\left( t,y\right) ,\ \ \ 0\leq y<\infty , \\ 
\\ 
r\left( 0,y\right) =0,\ \ \ r\left( t,-M\left( t\right) \right) =-\frac{\exp
\left( -\frac{\left( M\left( t\right) +z\right) ^{2}}{2t}\right) }{\sqrt{%
2\pi t}},\ \ \ r\left( t,y\rightarrow \infty \right) \rightarrow 0.%
\end{array}%
\end{equation*}

\subsection{Governing system of integral equations}

Using our standard approach, we obtain the following system of nonlinear
Volterra integral equations%
\begin{equation}
\begin{array}{c}
\nu \left( t\right) +\int_{0}^{t}\frac{\Theta \left( t,t^{\prime }\right)
\Xi \left( t,t^{\prime }\right) \nu \left( t^{\prime }\right) }{\sqrt{2\pi
\left( t-t^{\prime }\right) }}dt^{\prime }+H\left( t,t\Theta \left(
t,0\right) -z\right) =0, \\ 
\\ 
\mu \left( t\right) +\left( \frac{1}{\sqrt{2\pi t}}+\alpha \mu \left(
t\right) \right) \nu \left( t\right) + \\ 
\\ 
+\frac{1}{2}\int_{0}^{t}\frac{\Phi \left( t,t^{\prime }\right) +\Theta
^{2}\left( t,t^{\prime }\right) \Xi \left( t,t^{\prime }\right) \nu \left(
t^{\prime }\right) }{\sqrt{2\pi \left( t-t^{\prime }\right) }}dt^{\prime }+%
\frac{\left( t\Theta \left( t,0\right) -z\right) H\left( t,t\Theta \left(
t,0\right) -z\right) }{2t}=0,%
\end{array}
\label{Eq57}
\end{equation}%
where%
\begin{equation*}
\begin{array}{c}
\ \Theta \left( t,t^{\prime }\right) =\frac{\alpha \int_{t^{\prime }}^{t}\mu
\left( t^{\prime \prime }\right) dt^{\prime \prime }}{\left( t-t^{\prime
}\right) },\ \ \ \Xi \left( t,t^{\prime }\right) =e^{-\frac{\left(
t-t^{\prime }\right) \Theta ^{2}\left( t,t^{\prime }\right) }{2}},\ \ \ \Phi
\left( t,t^{\prime }\right) =\frac{\left( \nu \left( t\right) -\Xi \left(
t,t^{\prime }\right) \nu \left( t^{\prime }\right) \right) }{\left(
t-t^{\prime }\right) }, \\ 
\\ 
\Theta \left( t,t\right) =\alpha \mu \left( t\right) ,\ \ \ \Xi \left(
t,t\right) =1,\ \ \ \Phi \left( t,t\right) =\nu ^{\prime }\left( t\right) +%
\frac{1}{2}\alpha ^{2}\mu ^{2}\left( t\right) \nu \left( t\right) .%
\end{array}%
\end{equation*}

\subsection{Numerical solution}

In the spirit of Eq. (\ref{Eq24}), we get the following approximation for
Eqs (\ref{Eq57}) for $k>0$:%
\begin{equation*}
\begin{array}{c}
\nu _{k}+\frac{1}{\sqrt{2\pi }}\sum_{l=1}^{k}\left( P_{k,l}^{\left( 1\right)
}\nu _{l}+P_{k,l-1}^{\left( 1\right) }\nu _{l-1}\right) \Pi _{k,l}+\vartheta
_{k}=0, \\ 
\\ 
\mu _{k}+\left( \frac{1}{\sqrt{2\pi \Delta _{k,0}}}+\alpha \mu _{k}\right)
\nu _{k}+\frac{1}{2\sqrt{2\pi }}\sum_{l=1}^{k}\left( \Phi _{k,l}+\Phi
_{k,l-1}+P_{k,l}^{\left( 2\right) }\nu _{l}+P_{k,l-1}^{\left( 2\right) }\nu
_{l-1}\right) \Pi _{k,l}+\iota _{k}=0.%
\end{array}%
\end{equation*}%
Here and below we use the following notation%
\begin{equation*}
\begin{array}{c}
\Theta _{k,l}=\ \alpha \frac{\sum_{i=l+1}^{k}\left( \mu _{i}+\mu
_{i-1}\right) \Delta _{i,i-1}}{2\Delta _{k,l}},\ \ \ P_{k,l}^{\left(
i\right) }=\Theta _{k,l}^{i}e^{-\frac{\Delta _{k,l}\Theta _{k,l}^{2}}{2}},\
\ \ Q_{k,l}=P_{k,l}^{\left( 2\right) }-\frac{P_{k,l}^{\left( 0\right) }}{%
\Delta _{k,l}},\ \ \ \Phi _{k,l}=\frac{\nu _{k}-P_{k,l}^{\left( 0\right)
}\nu _{k-1}}{\Delta _{k,l}},\ \ \ k>l, \\ 
\\ 
\Theta _{k,k}=\ \alpha \mu _{k},\ \ \ P_{k,k}^{\left( i\right) }=\ \alpha
^{i}\mu _{k}^{i},\ \ \ Q_{k,k}\ \text{undefined},\ \ \ \Phi _{k,k}=\frac{\nu
_{k}-\nu _{k-1}}{\Delta _{k,k-1}}+\frac{1}{2}\alpha ^{2}\mu _{k}^{2}\nu
_{k},\  \\ 
\\ 
\vartheta _{k}=H\left( \Delta _{k,0},\Delta _{k,0}\Theta _{k,0}-z\right) ,\
\ \ \ \iota _{k}=\frac{\left( \Delta _{k,0}\Theta _{k,0}-z\right) \vartheta
_{k}}{2\Delta _{k,0}},\ \ \ k>0.%
\end{array}%
\end{equation*}

For $k=0$ we have: 
\begin{equation*}
\begin{array}{c}
\left( \nu _{0},\mu _{0}\right) =\left( 0,0\right) .%
\end{array}%
\end{equation*}

For $k=1$ we have:%
\begin{equation*}
\begin{array}{c}
\nu _{1}=-\frac{H\left( \Delta _{1,0},\frac{\Delta _{1,0}\alpha \mu _{1}}{2}%
-z\right) }{\left( 1+\sqrt{\frac{\Delta _{1,0}}{2\pi }}\alpha \mu
_{1}\right) }, \\ 
\\ 
\mu _{1}-\left( \frac{\left( \frac{1}{\sqrt{2\pi \Delta _{1,0}}}+\alpha \mu
_{1}+\frac{\alpha ^{2}\mu _{1}^{2}}{2\sqrt{2\pi \Delta _{1,0}}}\right) }{%
\left( 1+\sqrt{\frac{\Delta _{1,0}}{2\pi }}\alpha \mu _{1}\right) }-\frac{%
\left( \frac{\Delta _{1,0}\alpha \mu _{1}}{2}-z\right) }{2\Delta _{1,0}}%
\right) H\left( \Delta _{1,0},\frac{\Delta _{1,0}\alpha \mu _{1}}{2}%
-z\right) =0,%
\end{array}%
\end{equation*}%
where the nonlinear equation for $\mu _{1}$ has to be solved by the
Newton-Raphson method.

For $k>1$ we have%
\begin{equation*}
\begin{array}{c}
\left( 1+\sqrt{\frac{\Delta _{k,k-1}}{2\pi }}\alpha \mu _{k}\right) \nu _{k}+%
\sqrt{\frac{\Delta _{k,k-1}}{2\pi }}P_{k,k-1}^{\left( 1\right) }\nu _{k-1}
\\ 
\\ 
+\frac{1}{\sqrt{2\pi }}\sum_{l=1}^{k-1}\left( P_{k,l}^{\left( 1\right) }\nu
_{l}+P_{k,l-1}^{\left( 1\right) }\nu _{l-1}\right) \Pi _{k,l}+\vartheta
_{k}=0, \\ 
\\ 
\mu _{k}+\left( \frac{1}{\sqrt{2\pi \Delta _{k,0}}}+\alpha \mu _{k}+\frac{%
\alpha ^{2}\mu _{k}^{2}}{2\sqrt{2\pi \Delta _{k,k-1}}}+\frac{1}{2\sqrt{2\pi }%
}\sum_{l=1}^{k-1}\frac{\left( \Delta _{k,l}+\Delta _{k,l-1}\right) \Pi _{k,l}%
}{\Delta _{k,l}\Delta _{k,l-1}}\right) \nu _{k} \\ 
\\ 
+\frac{1}{2}\sqrt{\frac{\Delta _{k,k-1}}{2\pi }}Q_{k,k-1}\nu _{k-1}+\frac{1}{%
2\sqrt{2\pi }}\sum_{l=1}^{k-1}\left( Q_{k,l}\nu _{l}+Q_{k,l-1}\nu
_{l-1}\right) \Pi _{k,l}+\iota _{k}=0.%
\end{array}%
\end{equation*}%
Assuming that $\left( \nu _{1},\mu _{1}\right) ,\ldots ,\left( \nu
_{k-1},\mu _{k-1}\right) $ are known, we can express $\nu _{k}$ in terms of $%
\mu _{k}$:%
\begin{equation*}
\begin{array}{c}
\nu _{k}=-\frac{\left( \sqrt{\frac{\Delta _{k,k-1}}{2\pi }}P_{k,k-1}^{\left(
1\right) }\nu _{k-1}+\frac{1}{\sqrt{2\pi }}\sum_{l=1}^{k-1}\left(
P_{k,l}^{\left( 1\right) }\nu _{l}+P_{k,l-1}^{\left( 1\right) }\nu
_{l-1}\right) \Pi _{k,l}+\vartheta _{k}\right) }{\left( 1+\sqrt{\frac{\Delta
_{k,k-1}}{2\pi }}\alpha \mu _{k}\right) },%
\end{array}%
\end{equation*}%
and obtain a nonlinear equation for $\mu _{k}$:%
\begin{equation*}
\begin{array}{c}
\mu _{k}-\frac{\left( \frac{1}{\sqrt{2\pi \Delta _{k,0}}}+\alpha \mu _{k}+%
\frac{\alpha ^{2}\mu _{k}^{2}}{2\sqrt{2\pi \Delta _{k,k-1}}}+\frac{1}{2\sqrt{%
2\pi }}\sum_{l=1}^{k-1}\frac{\left( \Delta _{k,l}+\Delta _{k,l-1}\right) \Pi
_{k,l}}{\Delta _{k,l}\Delta _{k,l-1}}\right) }{\left( 1+\sqrt{\frac{\Delta
_{k,k-1}}{2\pi }}\alpha \mu _{k}\right) } \\ 
\\ 
\times \left( \sqrt{\frac{\Delta _{k,k-1}}{2\pi }}P_{k,k-1}^{\left( 1\right)
}\nu _{k-1}+\frac{1}{\sqrt{2\pi }}\sum_{l=1}^{k-1}\left( P_{k,l}^{\left(
1\right) }\nu _{l}+P_{k,l-1}^{\left( 1\right) }\nu _{l-1}\right) \Pi
_{k,l}+\vartheta _{k}\right) \\ 
\\ 
+\frac{\left( Q_{k,k-1}\Delta _{k,k-1}-1\right) \nu _{k-1}}{2\sqrt{2\pi
\Delta _{k,k-1}}}+\frac{1}{2\sqrt{2\pi }}\sum_{l=1}^{k-1}\left( Q_{k,l}\nu
_{l}+Q_{k,l-1}\nu _{l-1}\right) \Pi _{k,l}+\iota _{k}=0,%
\end{array}%
\end{equation*}%
which again is solved by the Newton-Raphson method.

\begin{figure}[tbp]
\begin{center}
{%
\includegraphics[width=1.0\textwidth]
{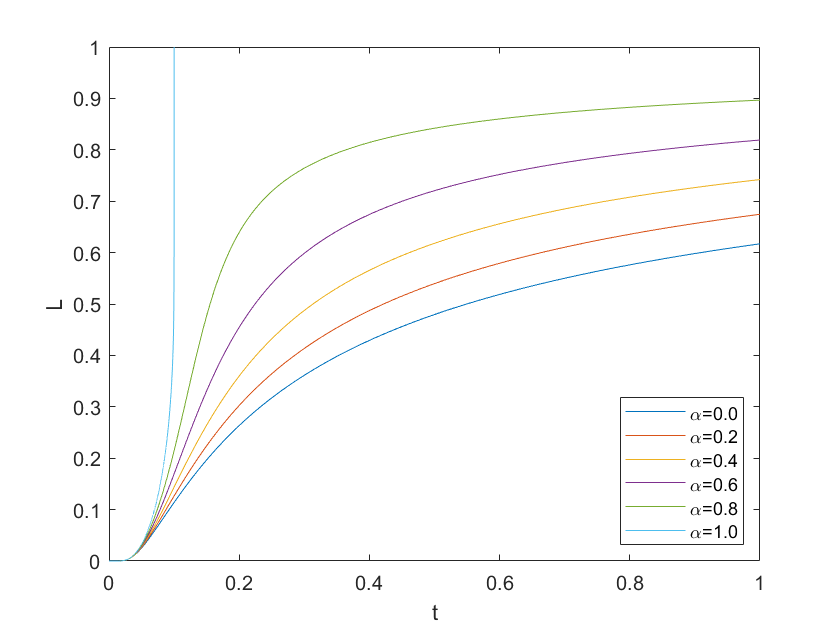}%
}
\end{center}
\par
\vspace{-10pt}
\caption{In this figure, we demonstrate the loss probability for the initial
position $z=0.5$ and several representative values of $\protect\alpha $,
which characterizes the strength of interbank interactions. }
\label{Fig4}
\end{figure}

In Figure \ref{Fig4} we show cumulative loss probability for several
representative values of $\alpha $.%
\begin{equation*}
\text{Figure \ref{Fig4} near here.}
\end{equation*}%
A striking feature of this figure is the "phase transition" occurring at $%
\alpha \approx 1.0$, when default after a finite time becomes inevitable. By
contrast, for $\alpha =0$, the default probability reaches unity only
asymptotically when $t\rightarrow \infty $.

\begin{figure}[tbp]
\begin{center}
\subfloat[]{\includegraphics[width=0.5\textwidth]
{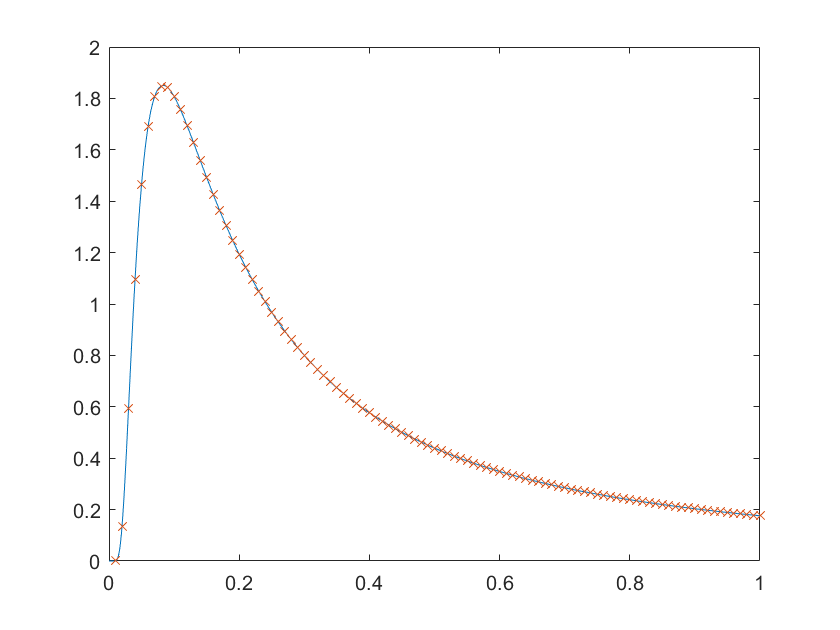}}
\\[0pt]
\subfloat[]{\includegraphics[width=0.5\textwidth]
{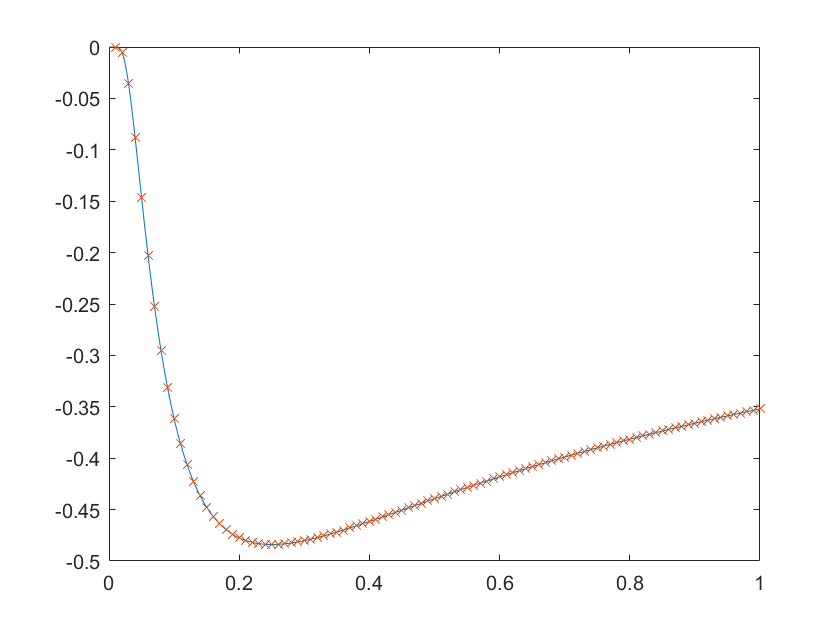}}
\\[0pt]
\subfloat[]{\includegraphics[width=0.5\textwidth]
{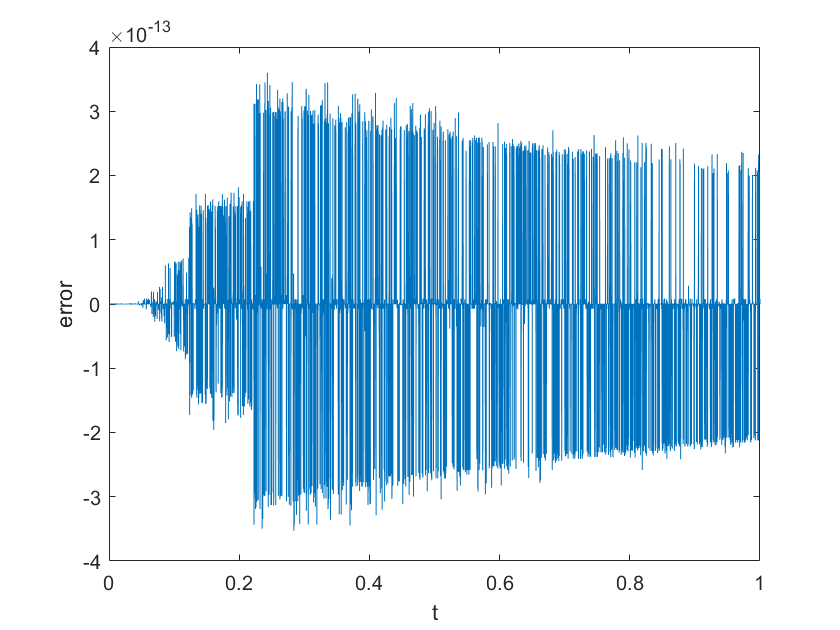}}
\end{center}
\par
\vspace{-10pt}
\caption{In Figures (a)-(b), we choose $z=0.5,$ $\protect\alpha =0,$ and we
show $\protect\mu \left( t\right) $ and $\protect\nu \left( t\right) $
calculated numerically and analytically. In Figure (c), we choose $z=0.5,$ $%
\protect\alpha =0.6,$ and we show the error generated by the Newton-Raphson
method.}
\label{Fig5}
\end{figure}

\begin{figure}[tbp]
\begin{center}
\subfloat[]{\includegraphics[width=0.5\textwidth]
{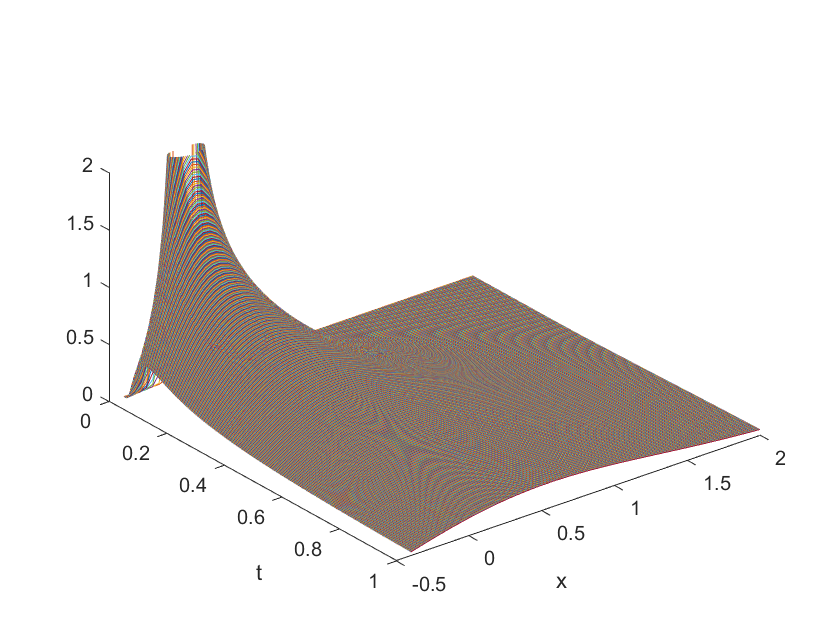}}
\subfloat[]{\includegraphics[width=0.5\textwidth]
{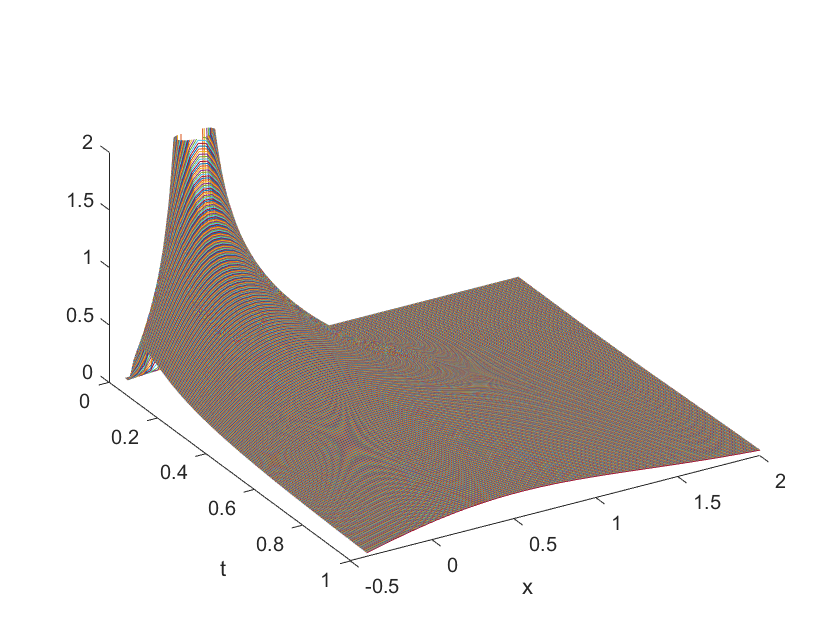}}
\\[0pt]
\subfloat[]{\includegraphics[width=0.5\textwidth]
{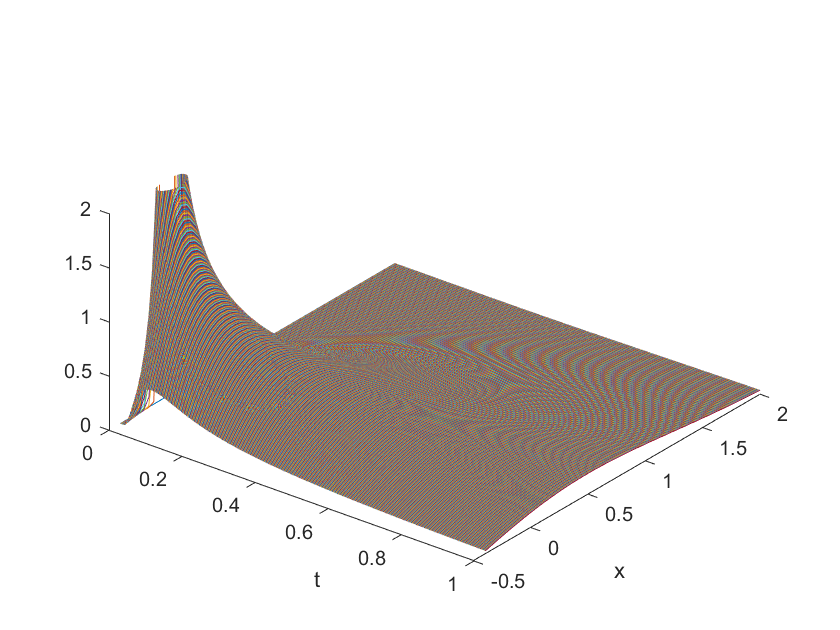}}%
\subfloat[]{\includegraphics[width=0.5\textwidth]
{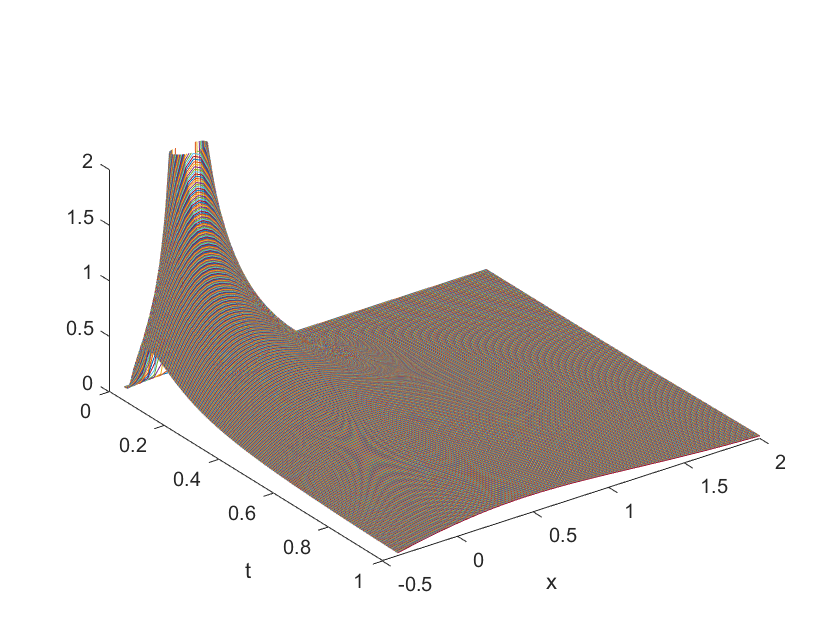}}%
\\[0pt]
\subfloat[]{\includegraphics[width=0.5\textwidth]
{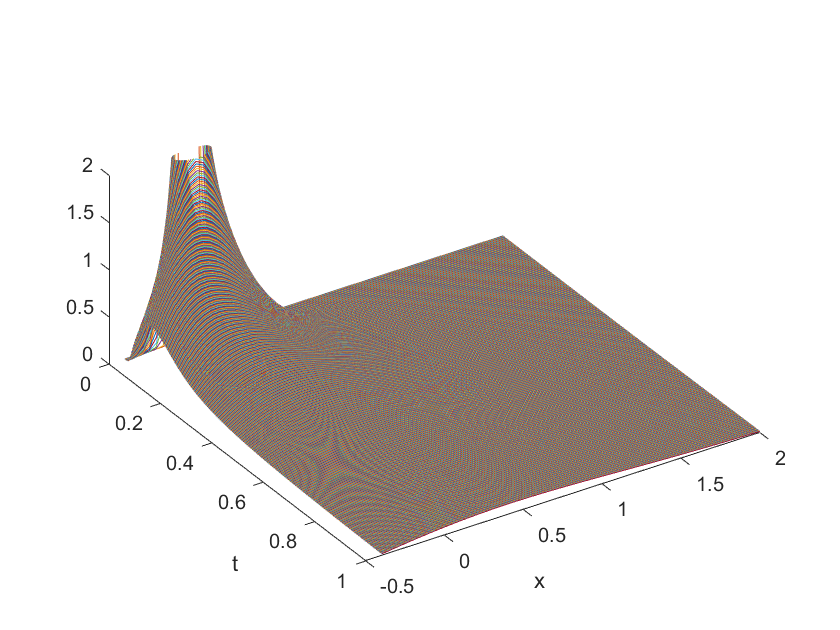}}%
\subfloat[]{\includegraphics[width=0.5\textwidth]
{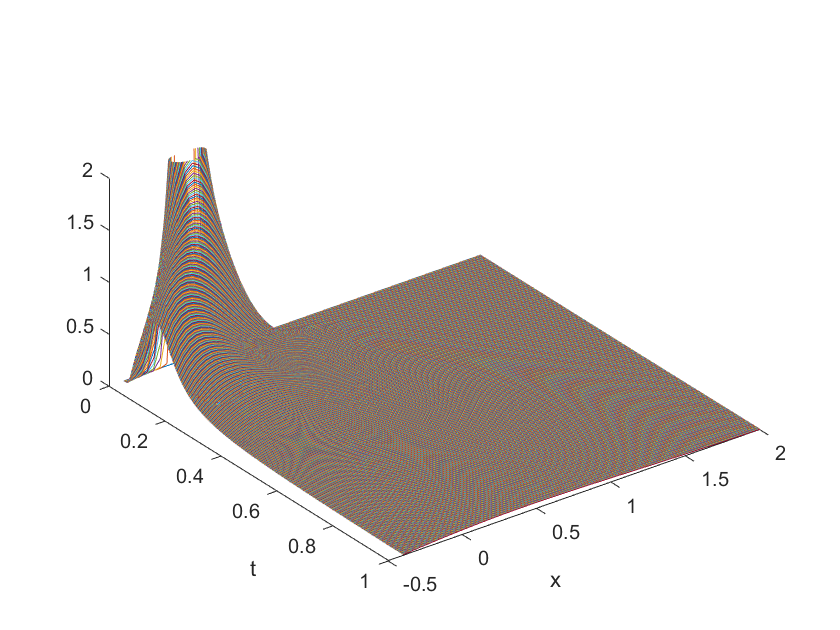}}
\end{center}
\par
\vspace{-10pt}
\caption{In Figures (a)-(f) we show the probability denstity function $%
p\left( t,x-z;z\right) $. We shift the domain down by $-z$ in order to make
comparison with the structural default model considered in Section \protect
\ref{Strudefa} more transparent. In Figures (a) and (b) we show analytical
and numerical results for $\protect\alpha =0$. In Figures (c)-(f) we show
numerical results for $\protect\alpha =0.2$, $0.4$, $0.6$, $0.8$,
respectively.}
\label{Fig6}
\end{figure}

We notice that for $\alpha =0$, $\mu \left( t\right) $, $\nu \left( t\right) 
$ can be calculated analytically. For benchmarking purposes, we compare
numerical and analytical results in Figure \ref{Fig5}, (a), (b). As usual,
the efficiency of the Newton-Raphson method, which is illustrated in Figure %
\ref{Fig5} (c) is nothing short of miraculous.%
\begin{equation*}
\text{Figure \ref{Fig5} near here.}
\end{equation*}%
In Figure \ref{Fig6} we represent shifted probability density surfaces $%
p\left( t,x-z;z\right) $ for representative values of $\alpha $ used in
Figure \ref{Fig4}. 
\begin{equation*}
\text{Figure \ref{Fig5} near here.}
\end{equation*}%
The shift is made in order to make the connection with Section \ref{Strudefa}
more transparent; after this shift all the processes start at $0$ and the
boundaries are given by $b=-0.5$.

\section{Hitting time probability distribution for an Ornstein-Uhlenbeck\
process\label{OU}}

\subsection{Preliminaries}

In a seminal paper, Fortet developed an original approach to calculating
probability distribution of the hitting time for a diffusion process, \cite%
{Fortet1943}. Fortet's equation can be viewed as a variant of the
Einstein-Smoluchowski equation, \cite{Einstein1905, Smoluchowski1906}. A
general overview can be found in \cite{Borodin2012, Breiman1967}.

Numerous attempts to find an analytical result for the Ornstein-Uhlenbeck
(OU) process have been made since 1998 when Leblanc and Scaillet first
derived an analytical formula, which contained a mistake, \cite{Leblanc1998}%
. Two years later, Leblanc \textit{et al.} published a correction on the
paper, \cite{Leblanc2000}; unfortunately, the correction was erroneous as
well, as was shown by \cite{Going2003}.

Several authors used the Laplace transform to find a formal semi-analytical
solution, \cite{Alili2005, Linetsky2004, Ricciardi1988}.

In this section, we use the EMHP to calculate the distribution of the
hitting time for an OU process. Our approach is semi-analytical and can
handle both constant and time-dependent parameters. It is worth noting that
the latter case cannot be solved using the Laplace transform method.
Additional information can be found in \cite{Lipton2020a}.

\subsection{Main equations}

To calculate the density $g\left( t,z\right) $ of the hitting time
probability distribution, we need to solve the following forward problem%
\begin{equation}
\begin{array}{c}
\frac{\partial }{\partial t}p\left( t,x;z\right) =p\left( t,x;z\right) +x%
\frac{\partial }{\partial x}p\left( t,x;z\right) +\frac{1}{2}\frac{\partial
^{2}}{\partial x^{2}}p\left( t,x;z\right) , \\ 
\\ 
p\left( 0,x;z\right) =\delta _{z}\left( x\right) ,\ \ \ p\left( t,b\left(
t\right) ;z\right) =0,\ \ \ p\left( t,x;z\rightarrow \infty \right)
\rightarrow 0.%
\end{array}
\label{Eq66}
\end{equation}%
This distribution is given by%
\begin{equation*}
g\left( t,z\right) =\frac{1}{2}\frac{\partial }{\partial x}p\left(
t,b;z\right) .
\end{equation*}

\subsection{Particular case, $b=0$}

Before solving the general problem via the EMHP, let us consider a
particular case of $b=0$. Green's function for the OU process in question
has the form%
\begin{equation*}
\begin{array}{c}
G\left( t,x;z\right) =e^{t}H\left( \eta \left( t\right) ,e^{t}x-z\right) ,%
\end{array}%
\end{equation*}%
where 
\begin{equation*}
\begin{array}{c}
\eta \left( t\right) =\frac{e^{2t}-1}{2}=e^{t}\sinh \left( t\right) .%
\end{array}%
\end{equation*}%
Since $b=0$, the method of images works, so that%
\begin{equation*}
\begin{array}{c}
p\left( t,x;z\right) =e^{t}H\left( \eta \left( t\right) ,e^{t}x-z\right)
-e^{t}H\left( \eta \left( t\right) ,e^{t}x+z\right) , \\ 
\\ 
g\left( t\right) =\frac{1}{2}\frac{\partial }{\partial x}p\left( t,0\right) =%
\frac{ze^{2t}H\left( \eta \left( t\right) ,-z\right) }{\eta \left( t\right) }%
, \\ 
\\ 
G\left( t\right) =\int_{0}^{t}g\left( t^{\prime }\right) dt^{\prime
}=2N\left( -\frac{z}{\sqrt{\eta \left( t\right) }}\right) .%
\end{array}%
\end{equation*}%
This result is useful for benchmarking purposes.

\subsection{General case}

To be concrete, consider the case $z>b(0)$. {We wish to transform the IBVP (%
\ref{Eq66}) } into the standard IBVP for a heat equation with a moving
boundary. To this end, we introduce new independent and dependent variables
as follows:%
\begin{equation}
\begin{array}{c}
q(\tau ,\xi )=e^{-t}p(t,x),\quad \tau =\eta \left( t\right) ,\quad \xi
=e^{t}x, \\ 
\\ 
p\left( t,x\right) =\sqrt{1+2\tau }q\left( \tau ,\xi \right) ,\ \ \ t=\ln
\left( \sqrt{1+2\tau }\right) ,\ \ \ x=\frac{\xi }{\sqrt{1+2\tau }},%
\end{array}
\label{Eq71}
\end{equation}%
and get the IBVP of the form%
\begin{equation*}
\begin{array}{c}
\frac{\partial }{\partial \tau }q\left( \tau ,\xi \right) =\frac{1}{2}\frac{%
\partial ^{2}}{\partial \xi ^{2}}q\left( \tau ,\xi \right) ,\ \ \ \ \ \beta
\left( \tau \right) \leq \xi <\infty , \\ 
\\ 
q\left( 0,\xi \right) =\delta _{z}\left( \xi \right) ,\ \ \ q\left( \tau
,\beta \left( \tau \right) \right) =0,\ \ \ q\left( \tau ,\xi \rightarrow
\infty \right) \rightarrow 0.%
\end{array}%
\end{equation*}%
Here 
\begin{equation*}
\begin{array}{c}
\beta \left( \tau \right) =\sqrt{1+2\tau }\tilde{b}(\ln \left( \sqrt{1+2\tau 
}\right) ).%
\end{array}%
\end{equation*}

\subsection{The governing system of integral equations}

The corresponding system of Volterra integral equations has the form%
\begin{equation}
\begin{array}{c}
\nu \left( \tau \right) +\int_{0}^{\tau }\frac{\Theta \left( \tau ,\tau
^{\prime }\right) \Xi \left( \tau ,\tau ^{\prime }\right) \nu \left( \tau
^{\prime }\right) }{\sqrt{2\pi \left( \tau -\tau ^{\prime }\right) }}d\tau
^{\prime }+H\left( \tau ,\beta \left( \tau \right) -z\right) =0, \\ 
\\ 
\mu \left( \tau \right) +\left( \frac{1}{\sqrt{2\pi \tau }}+\beta ^{\prime
}\left( \tau \right) \right) \nu \left( \tau \right) +\frac{1}{2}%
\int_{0}^{\tau }\frac{\Phi \left( \tau ,\tau ^{\prime }\right) +\Theta
^{2}\left( \tau ,\tau ^{\prime }\right) \Xi \left( \tau ,\tau ^{\prime
}\right) \nu \left( \tau ^{\prime }\right) }{\sqrt{2\pi \left( \tau -\tau
^{\prime }\right) }}d\tau ^{\prime }+\frac{\left( \beta \left( \tau \right)
-z\right) H\left( \tau ,\beta \left( \tau \right) -z\right) }{2\tau }=0,%
\end{array}
\label{Eq74}
\end{equation}%
where%
\begin{equation*}
\begin{array}{c}
\mu \left( \tau \right) =\left( 1+2\tau \right) g\left( \ln \left( \sqrt{%
1+2\tau }\right) \right) .%
\end{array}%
\end{equation*}%
This system is linear, so that $\mu \left( \tau \right) $ is expressed in
terms of $\nu \left( \tau \right) $ directly and there is no need to use the
Newton-Raphson method.

\subsection{Flat boundary}

Assuming that the boundary is flat, we can simplify Eqs (\ref{Eq74})
somewhat. We notice that 
\begin{equation*}
\begin{array}{c}
\frac{\beta \left( \tau \right) -\beta \left( \tau ^{\prime }\right) }{\tau
-\tau ^{\prime }}=\frac{2b}{\sqrt{1+2\tau }+\sqrt{1+2\tau ^{\prime }}},%
\end{array}%
\end{equation*}%
introduce%
\begin{equation*}
\begin{array}{c}
\theta =\sqrt{1+2\tau }-1,\theta ^{\prime }=\sqrt{1+2\tau ^{\prime }}%
-1,0\leq \theta ^{\prime }\leq \theta <\infty ,%
\end{array}%
\end{equation*}%
and write the first equation (\ref{Eq74}) in the form 
\begin{equation}
\begin{array}{c}
\nu \left( \theta \right) +\frac{2b}{\sqrt{\pi }}\int_{0}^{\theta }\frac{%
\exp \left( -\frac{b^{2}\left( \theta -\theta ^{\prime }\right) }{\left(
2+\theta +\theta ^{\prime }\right) }\right) \left( 1+\theta ^{\prime
}\right) \nu \left( \theta ^{\prime }\right) }{\sqrt{\left( 2+\theta +\theta
^{\prime }\right) ^{3}\left( \theta -\theta ^{\prime }\right) }}\ d\theta
^{\prime }+\frac{e^{-\frac{\left( \left( 1+\theta \right) b-z\right) ^{2}}{%
\left( \left( 1+\theta \right) ^{2}-1\right) }}}{\sqrt{\pi \left( \left(
1+\theta \right) ^{2}-1\right) }}=0.%
\end{array}
\label{Eq77}
\end{equation}%
Provided that $\nu \left( \theta \right) $ is known, we can represent $%
g\left( t\right) $ is the form 
\begin{equation*}
\begin{array}{c}
g\left( t\right) =-\frac{\left( e^{t}b-z\right) \exp \left( -\frac{\left(
e^{t}b-z\right) ^{2}}{\left( e^{2t}-1\right) }+2t\right) }{\sqrt{\pi \left(
e^{2t}-1\right) ^{3}}}-\left( e^{t}b+\frac{e^{2t}}{\sqrt{\pi \left(
e^{2t}-1\right) }}\right) \nu \left( t\right) \\ 
\\ 
+\frac{1}{\sqrt{\pi }}e^{2t}\int_{0}^{\theta }\frac{\left( \left( 1-2b^{2}%
\frac{\left( \theta -\theta ^{\prime }\right) }{\left( 2+\theta +\theta
^{\prime }\right) }\right) \exp \left( -b^{2}\frac{\left( \theta -\theta
^{\prime }\right) }{\left( 2+\theta +\theta ^{\prime }\right) }\right) \nu
\left( \theta ^{\prime }\right) -\nu \left( \theta \right) \right) \left(
1+\theta ^{\prime }\right) }{\sqrt{\left( 2+\theta +\theta ^{\prime }\right)
^{3}\left( \theta -\theta ^{\prime }\right) ^{3}}}d\theta ^{\prime }.%
\end{array}%
\end{equation*}

It is worth noting that the analytical solution is available in two cases:
(A) when\ $b=0$ the solution can be found by using the method of images; (B)
when $b(t)=Ae^{-t}+Be^{t}$ the boundary transforms into the linear boundary $%
2B\tau +A+B$, which can be treated by the method of images as well.

\begin{figure}[tbp]
\begin{center}
\subfloat[]{\includegraphics[width=0.5\textwidth]
{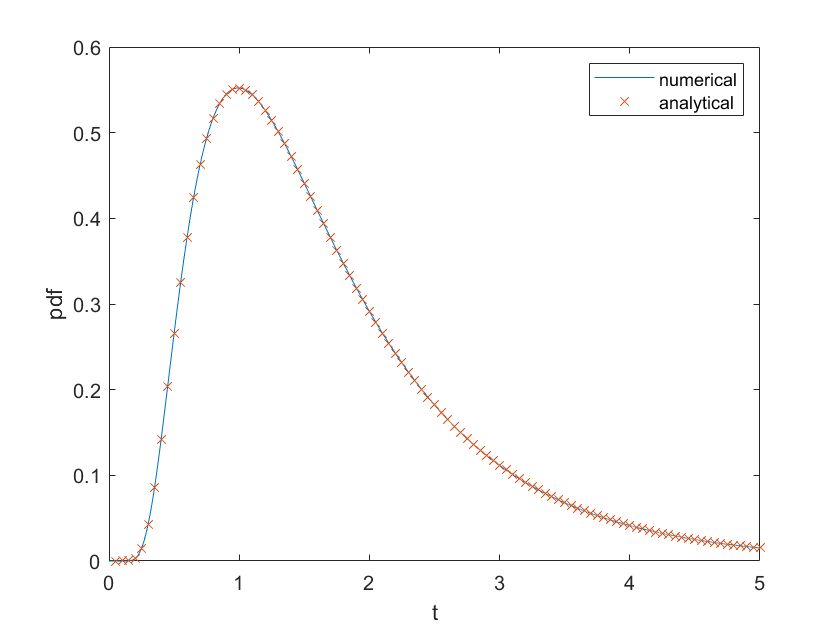}}
\subfloat[]{\includegraphics[width=0.5\textwidth]
{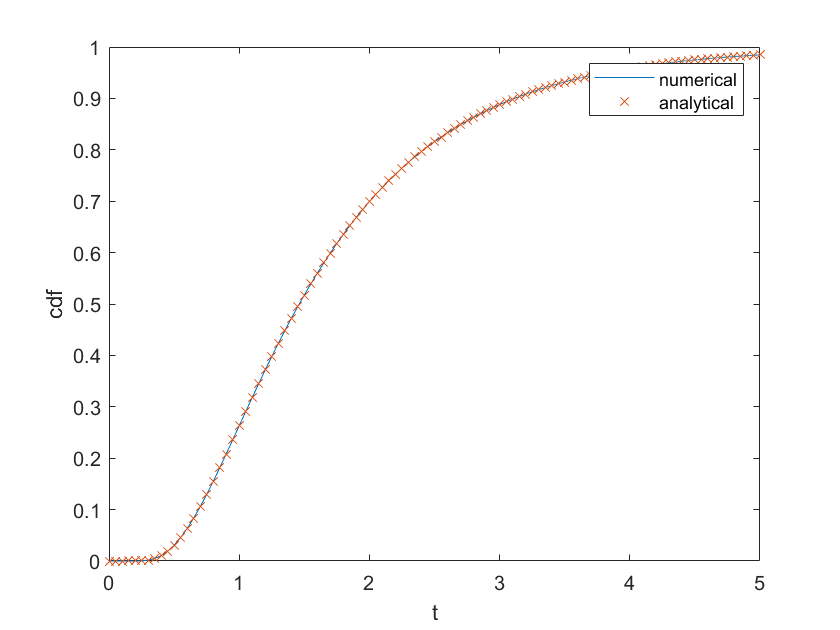}}
\\[0pt]
\subfloat[]{\includegraphics[width=0.5\textwidth]
{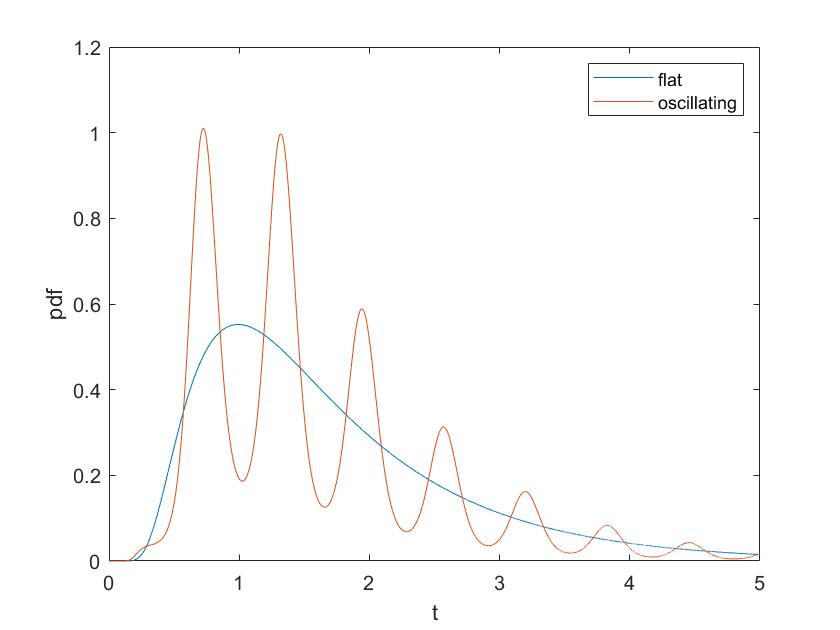}}%
\subfloat[]{\includegraphics[width=0.5\textwidth]
{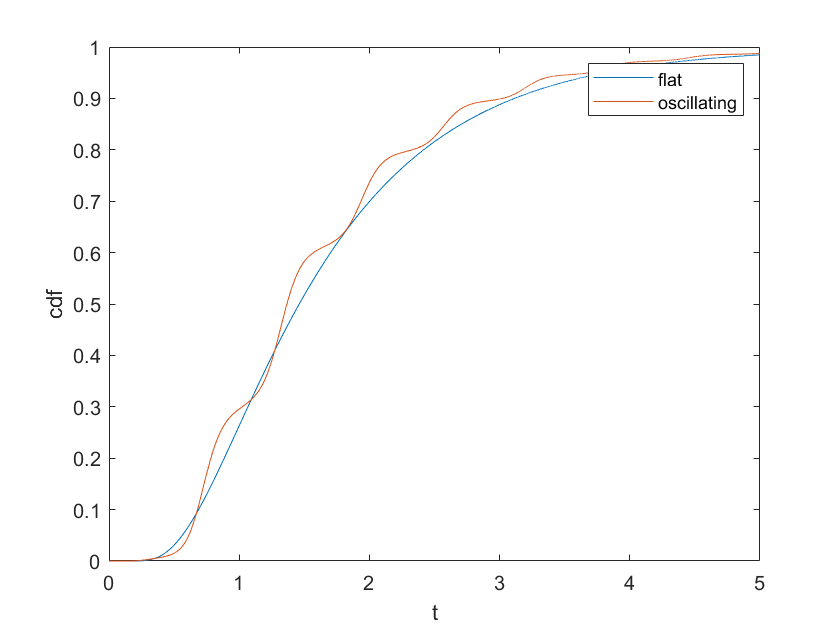}}%
\\[0pt]
\subfloat[]{\includegraphics[width=0.5\textwidth]
{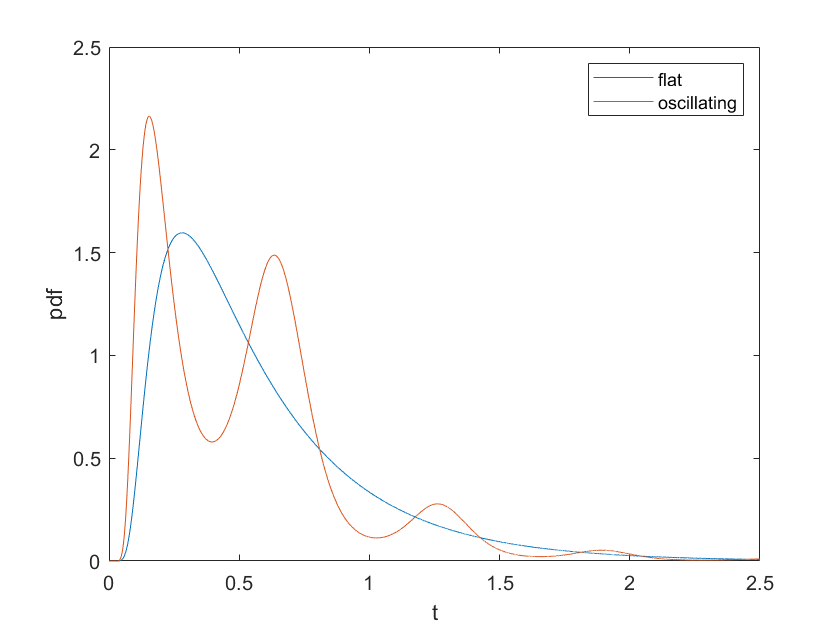}}%
\subfloat[]{\includegraphics[width=0.5\textwidth]
{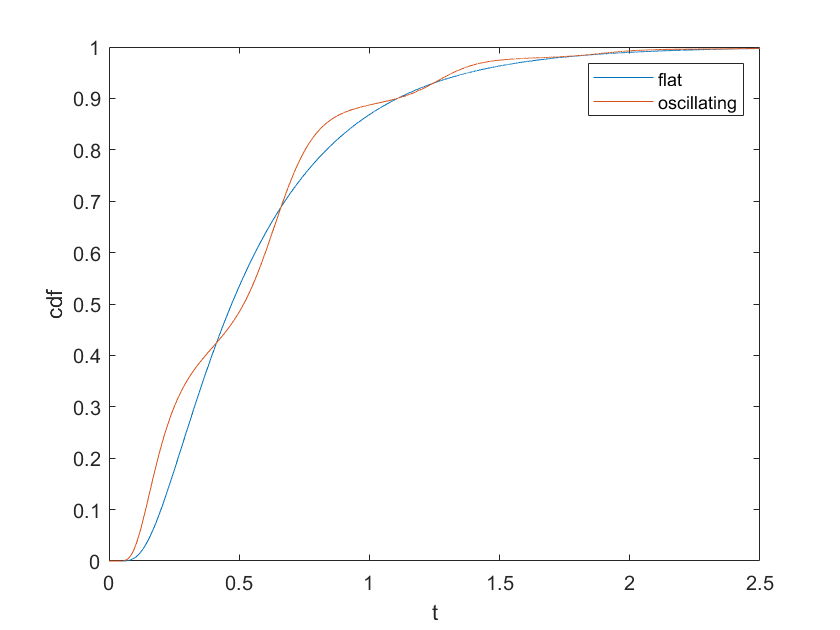}}
\end{center}
\par
\vspace{-10pt}
\caption{In Figures (a)-(f), we show the pdf and cdf for the hitting time
probability distribution. In Figures (a), (b) $z=2$, $b\left( t\right) =0$,
so that both numerical and analytical expressions are available. These
expressions are in perfect agreement. In Figures (c), (d) $z=2$, $b\left(
t\right) =0$ and $b\left( t\right) =0.2\sin \left( 10.0t\right) $. In
Figures (e), (f) $z=2$, $b\left( t\right) =1.0$ and $b\left( t\right)
=1.0+0.2\sin \left( 10.0t\right) $. Variations in the pdf caused by the
barrier undulations are astonishingly profound.}
\label{Fig7}
\end{figure}

We show the probability density function (pdf) and the cumulative density
function (cdf) for the hitting time in Figure \ref{Fig7}. It is interesting
to note that the undulation of the boundary causes considerable variations
in the pdfs, which are naturally less pronounced for the corresponding cdfs.%
\begin{equation*}
\text{Figure \ref{Fig7} near here.}
\end{equation*}

\subsection{Abel integral equation}

Consider Eq. (\ref{Eq77}), which we got for the standard OU process. For
small values of $\theta $, this equation can be approximated by an Abel
integral equation of the second kind.%
\begin{equation*}
\begin{array}{c}
\nu \left( \theta \right) +\frac{b}{\sqrt{2\pi }}\int_{0}^{\theta }\frac{1}{%
\sqrt{\theta -\theta ^{\prime }}}\nu \left( \theta ^{\prime }\right) d\theta
^{\prime }+H\left( \theta ,b-z\right) =0.%
\end{array}%
\end{equation*}%
This equation can be solved analytically using direct - inverse Laplace
transforms. The direct Laplace transform yields 
\begin{equation*}
\begin{array}{c}
\bar{\nu}(\Lambda )+b\frac{\bar{\nu}(\Lambda )}{\sqrt{2\Lambda }}+\frac{e^{-%
\sqrt{2\Lambda }(z-b)}}{\sqrt{2\Lambda }}=0.%
\end{array}%
\end{equation*}%
Then, $\bar{\nu}(\Lambda )$ can be expressed as 
\begin{equation*}
\begin{array}{c}
\bar{\nu}(\Lambda )=-\frac{e^{-\sqrt{2\Lambda }(z-b)}}{\sqrt{2\Lambda }+b}.%
\end{array}%
\end{equation*}%
Taking the inverse Laplace transform, we get the final expression for $\nu
(\theta )$ 
\begin{equation*}
\begin{array}{c}
\nu (\theta )=be^{\frac{b^{2}}{2}\theta +b(z-b)}N\left( -\frac{b\theta +z-b}{%
\sqrt{\theta }}\right) -\frac{\exp \left( -\frac{\left( b-z\right) ^{2}}{%
2\theta }\right) }{\sqrt{2\pi \theta }}.%
\end{array}%
\end{equation*}

Alternatively, one can represent an analytical solution of an Abel equation 
\begin{equation*}
\begin{array}{c}
y(t)+\xi \int_{0}^{t}\frac{y(s)ds}{\sqrt{t-s}}=f(t).%
\end{array}%
\end{equation*}%
in the form 
\begin{equation*}
\begin{array}{c}
y(t)=F(t)+\pi \xi ^{2}\int_{0}^{t}\exp [\pi \xi ^{2}(t-s)]F(s)\,ds,%
\end{array}%
\end{equation*}%
where 
\begin{equation*}
\begin{array}{c}
F(t)=f(t)-\xi \int_{0}^{t}\frac{f(s)\,ds}{\sqrt{t-s}},%
\end{array}%
\end{equation*}%
see \cite{Polyanin1998}.

Abel equations naturally arise in many financial mathematics situations,
mainly, when fractional differentiation is involved, see, e.g., \cite%
{Andersen2013}.

\section{The supercooled Stefan problem}

The Stefan problem is of great theoretical and practical interest, see,
e.g., \cite{Kamenomostskaya1961, Rubinstein1971, Delarue2019} and references
therein. The classical Stefan problem studies the evolving boundary between
the two phases of the same medium, such as ice and water. Thus, this problem
boils down to solving the heat equation with a free boundary, which is
determined by a matching condition. The main equations for the supercooled
Stefan problem, are very similar to the mean-field banking equations:%
\begin{equation*}
\begin{array}{c}
\frac{\partial }{\partial t}p\left( t,x\right) =\frac{1}{2}\frac{\partial
^{2}}{\partial x^{2}}p\left( t,x\right) ,\ \ \ b\left( t\right) \leq
x<\infty , \\ 
\\ 
p\left( 0,x\right) =\delta _{z}\left( x\right) ,\ \ \ \ p\left( t,b\left(
t\right) \right) =0,\ \ \ p\left( t,X\rightarrow \infty \right) \rightarrow
0,\ 
\end{array}%
\end{equation*}%
where $p$ is the negative temperature profile, and $b$ is the liquid-solid
boundary. The location of the boundary is determined by the matching
condition%
\begin{equation*}
\begin{array}{c}
\frac{d}{dt}b\left( t\right) =\frac{\alpha }{2}\frac{\partial }{\partial x}%
p\left( t,x\right) .%
\end{array}%
\end{equation*}

As usual, we represent $p$ as $p=H+r$, where $r$ solves the following IBVP%
\begin{equation*}
\begin{array}{c}
\frac{\partial }{\partial t}r\left( t,x\right) =\frac{1}{2}\frac{\partial
^{2}}{\partial x^{2}}r\left( t,x\right) ,\ \ \ b\left( t\right) \leq
x<\infty , \\ 
\\ 
r\left( 0,x\right) =0,\ \ \ \ r\left( t,b\left( t\right) \right) =-H\left(
t,b\left( t\right) -z\right) ,\ \ r\left( t,X\rightarrow \infty \right)
\rightarrow 0,%
\end{array}%
\end{equation*}%
By using Eq. (\ref{Eq16}) we get the following system of coupled Volterra
equations:%
\begin{equation}
\begin{array}{c}
\nu \left( t^{\prime }\right) +\int_{0}^{t}\frac{\Theta \left( t,t^{\prime
}\right) \Xi \left( t,t^{\prime }\right) \nu \left( t^{\prime }\right) }{%
\sqrt{2\pi \left( t-t^{\prime }\right) }}dt^{\prime }+H\left( t,b\left(
t\right) -z\right) =0, \\ 
\\ 
b\left( t\right) +\frac{\alpha }{2}\int_{0}^{t}\frac{\Xi \left( t,t^{\prime
}\right) \nu \left( t^{\prime }\right) }{\sqrt{2\pi \left( t-t^{\prime
}\right) }}dt^{\prime }=0.%
\end{array}
\label{Eq86c}
\end{equation}%
where%
\begin{equation*}
\begin{array}{c}
\Theta \left( t,t^{\prime }\right) =\frac{b\left( t\right) -b\left(
t^{\prime }\right) }{\left( t-t^{\prime }\right) },\ \ \ \Xi \left(
t,t^{\prime }\right) =e^{-\frac{\left( t-t^{\prime }\right) \Theta
^{2}\left( t,t^{\prime }\right) }{2}},\ \ \ \ \Theta \left( t,t\right) =%
\frac{db\left( t\right) }{dt},\ \ \ \ \Xi \left( t,t\right) =1.%
\end{array}%
\end{equation*}%
System of integral equations (\ref{Eq86c}) is very similar to system (\ref%
{Eq36}) and can be solved by the same token.

\begin{figure}[tbp]
\begin{center}
{%
\includegraphics[width=1.0\textwidth]
{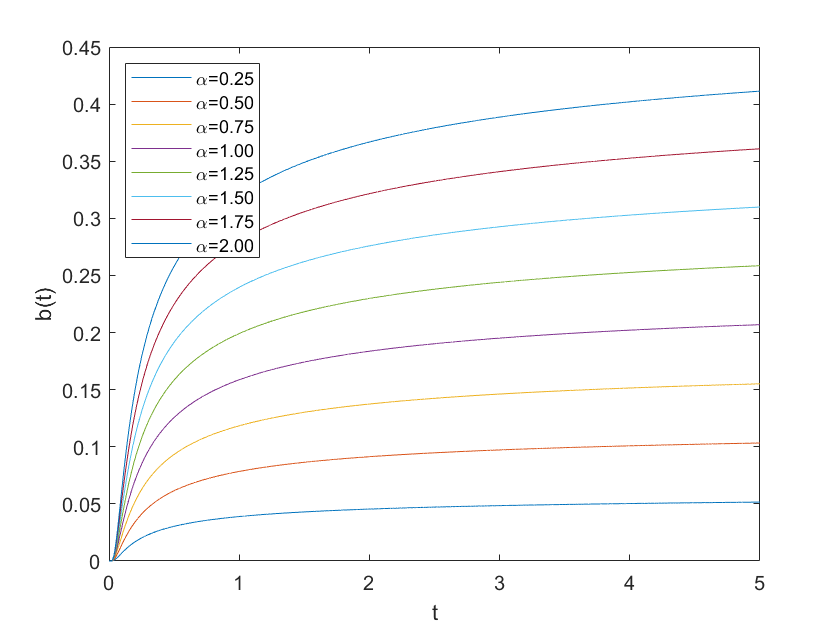}%
}
\end{center}
\par
\vspace{-10pt}
\caption{In this figure, we show the solid-liquid boundaries b(t) for
several representative values of $\protect\alpha $. }
\label{Fig8}
\end{figure}

In Figure \ref{Fig8} we show $b\left( t\right) $ for several representative
values of $\alpha $.%
\begin{equation*}
\text{Figure \ref{Fig8} near here.}
\end{equation*}

In this section, we deal with one of the rare instances when financial
mathematics results can be successfully used in the broader applied
mathematics context rather than the other way around.

\section{The integrate-and-fire neuron excitation model}

\subsection{Governing equations}

We briefly describe the famous integrate-and-fire model in neuroscience,
see, e.g., \cite{Lewis2003, Ostojic2009, Carrillo2013} . The neuron
excitation problem has the form:$\ \ $%
\begin{equation}
\begin{array}{c}
\frac{\partial }{\partial t}p\left( t,x\right) =\frac{\partial }{\partial x}%
\left( \left( x-\mu \left( t\right) \right) p\left( t,x\right) \right) +%
\frac{1}{2}\frac{\partial ^{2}}{\partial x^{2}}p\left( t,x\right) +\lambda
\left( t\right) \delta _{X_{0}}\left( x\right) ,\ \ -\infty <x\leq 0, \\ 
\\ 
p\left( 0,x\right) =p_{0}\left( x\right) ,\ \ \ p\left( t,-\infty \right)
=0,\ \ \ \ \ p\left( t,0\right) =0,\  \\ 
\\ 
X_{0}<0,\ \ \ \ \lambda \left( t\right) =-\frac{1}{2}\frac{\partial }{%
\partial x}p\left( t,0\right) ,\ \ \ \ \ \mu \left( t\right)
=m_{0}+m_{1}\lambda \left( t\right) ,%
\end{array}
\label{Eq87}
\end{equation}%
where $p\left( t,x\right) >0$ is the probability density of finding neurons
at a voltage $x$. Without loss of generality, we choose%
\begin{equation*}
\begin{array}{c}
p_{0}\left( x\right) =\delta _{\xi }\left( x\right) ,\ \ \ \xi <0.%
\end{array}%
\end{equation*}

Eqs (\ref{Eq87}) preserve probability in the sense that%
\begin{equation*}
\begin{array}{c}
\frac{d}{dt}\int_{-\infty }^{0}p\left( t,x\right) dx=0.%
\end{array}%
\end{equation*}%
Indeed, integration of the main equation yields:%
\begin{equation*}
\begin{array}{c}
\frac{d}{dt}\int_{-\infty }^{0}p\left( t,x\right) dx=\int_{-\infty }^{0}%
\frac{\partial }{\partial t}p\left( t,x\right) dx \\ 
\\ 
=\int_{-\infty }^{0}\left( \frac{\partial }{\partial x}\left( \left( x-\mu
\left( t\right) \right) p\left( t,x\right) \right) +\frac{1}{2}\frac{%
\partial ^{2}}{\partial x^{2}}p\left( t,x\right) +\lambda \left( t\right)
\delta _{X_{0}}\left( x\right) \right) dx \\ 
\\ 
=\frac{1}{2}\frac{\partial }{\partial x}p\left( t,0\right) +\lambda \left(
t\right) =0.%
\end{array}%
\end{equation*}

\subsection{The stationary problem}

Because the integrate-and-fire equations are probability-preserving, there
exists a stationary solution, which solves the time-independent
Fokker--Planck problem: 
\begin{equation*}
\begin{array}{c}
0=\frac{\partial }{\partial x}\left( \left( x-\mu \right) p\left( x\right)
\right) +\frac{1}{2}\frac{\partial ^{2}}{\partial x^{2}}p\left( x\right)
+\lambda \delta _{X_{0}}\left( x\right) , \\ 
\\ 
p\left( -\infty \right) =0,\ \ \ \ \ p\left( 0\right) =0,\ \ \ -\infty
<x\leq 0, \\ 
\\ 
\lambda =-\frac{1}{2}\frac{\partial }{\partial x}p\left( 0\right) ,\ \ \ \ \
\mu =m_{0}+m_{1}\lambda .%
\end{array}%
\end{equation*}%
We represent $p\left( x\right) $ in the form%
\begin{equation*}
\begin{array}{c}
p\left( x\right) =p^{<}\left( x\right) \left( 1-\Theta \left( x-X_{0}\right)
\right) +p^{>}\left( x\right) \Theta \left( x-X_{0}\right) ,%
\end{array}%
\end{equation*}%
where $\Theta \left( .\right) $ is the Heaviside function, and notice that%
\begin{equation}
\begin{array}{c}
p^{<}\left( X_{0}\right) =p^{>}\left( X_{0}\right) \equiv \nu , \\ 
\\ 
\frac{1}{2}\left( \ \frac{\partial }{\partial x}p^{>}\left( X_{0}\right) -\ 
\frac{\partial }{\partial x}p^{<}\left( X_{0}\right) \right) =-\lambda ,%
\end{array}
\label{Eq94}
\end{equation}%
where $\nu ,\lambda $ are unknown constants, which have to be determined as
part of the solution. In view of the boundary conditions, it is clear that%
\begin{equation*}
\begin{array}{c}
\left( x-\mu \right) p^{>}\left( x\right) +\frac{1}{2}\frac{\partial }{%
\partial x}p^{>}\left( x\right) =c^{>}\equiv -\lambda , \\ 
\\ 
\left( x-\mu \right) p^{<}\left( x\right) +\frac{1}{2}\frac{\partial }{%
\partial x}p^{<}\left( x\right) =c^{<}\equiv 0.%
\end{array}%
\end{equation*}%
Moreover, since $p$ is continuous at $x=X_{0}$ the second matching condition
(\ref{Eq94}) is satisfied automatically.

The method of separation of variables yields%
\begin{equation*}
\begin{array}{c}
p^{<}\left( x\right) =\nu e^{\left( X_{0}-\mu \right) ^{2}-\left( x-\mu
\right) ^{2}}.%
\end{array}%
\end{equation*}%
while the method of variation of constants yields%
\begin{equation*}
\begin{array}{c}
p^{>}\left( x\right) =2\lambda \left( e^{\mu ^{2}-\left( x-\mu \right)
^{2}}D\left( -\mu \right) -D\left( x-\mu \right) \right) ,%
\end{array}%
\end{equation*}%
where $D\left( .\right) $ is Dawson's integral,%
\begin{equation*}
\begin{array}{c}
D\left( x\right) =e^{-x^{2}}\int_{0}^{x}e^{y^{2}}dy.%
\end{array}%
\end{equation*}%
Thus,%
\begin{equation*}
\begin{array}{c}
\nu =2\lambda \left( e^{\mu ^{2}-\left( X_{0}-\mu \right) ^{2}}D\left( -\mu
\right) -D\left( X_{0}-\mu \right) \right) ,%
\end{array}%
\end{equation*}%
and%
\begin{equation*}
\begin{array}{c}
p^{<}\left( x\right) =2\lambda \left( e^{\mu ^{2}}D\left( -\mu \right)
-e^{\left( X_{0}-\mu \right) ^{2}}D\left( X_{0}-\mu \right) \right)
e^{-\left( x-\mu \right) ^{2}}.%
\end{array}%
\end{equation*}

At the same time, in the stationary case, the probability density $p\left(
x\right) $ has to integrate to unity:%
\begin{equation*}
\begin{array}{c}
\int_{-\infty }^{0}p\left( x\right) dx=\int_{-\infty }^{X_{0}}p^{<}\left(
x\right) dx+\int_{X_{0}}^{0}p^{>}\left( x\right) dx=1,%
\end{array}%
\end{equation*}%
which is a nonlinear equation for $\lambda $, because both $\mu $ and $\nu $
are known functions of $\lambda $. Once this equation is solved numerically,
the entire profile is determined. It is worth noting that the integral $%
\int_{-\infty }^{X_{0}}p^{<}\left( x\right) dx$ can be computed analytically:%
\begin{equation*}
\begin{array}{c}
\int_{-\infty }^{X_{0}}p^{<}\left( x\right) dx=\nu \int_{-\infty
}^{X_{0}}e^{\left( X_{0}-\mu \right) ^{2}-\left( x-\mu \right) ^{2}}dx=\sqrt{%
\pi }\nu e^{\left( X_{0}-\mu \right) ^{2}}N\left( \sqrt{2}\left( X_{0}-\mu
\right) \right) ,%
\end{array}%
\end{equation*}%
while the second integral $\int_{X_{0}}^{0}p^{>}\left( x\right) dx$ can be
split into two parts, the first of which can be computed analytically, and
the second one has to be computed numerically:%
\begin{equation*}
\begin{array}{c}
\int_{X_{0}}^{0}2\lambda \left( e^{-x\left( x-2\mu \right) }D\left( -\mu
\right) -D\left( x-\mu \right) \right) dx \\ 
\\ 
=2\lambda \left( \sqrt{\pi }e^{\mu ^{2}}\left( N\left( -\sqrt{2}\mu \right)
-N\left( \sqrt{2}\left( X_{0}-\mu \right) \right) \right) D\left( -\mu
\right) -\int_{X_{0}-\mu }^{-\mu }D\left( x\right) dx\right) .%
\end{array}%
\end{equation*}%
Thus, the corresponding nonlinear equation for $\lambda $ can be written as%
\begin{equation*}
\begin{array}{c}
\sqrt{\pi }\left( e^{\mu ^{2}}N\left( -\sqrt{2}\mu \right) D\left( -\mu
\right) -e^{\left( X_{0}-\mu \right) ^{2}}N\left( \sqrt{2}\left( X_{0}-\mu
\right) \right) D\left( X_{0}-\mu \right) \right) -\int_{X_{0}-\mu }^{-\mu
}D\left( x\right) dx-\frac{1}{2\lambda }=0.%
\end{array}%
\end{equation*}

\begin{figure}[tbp]
\begin{center}
\subfloat[]{\includegraphics[width=0.5\textwidth]
{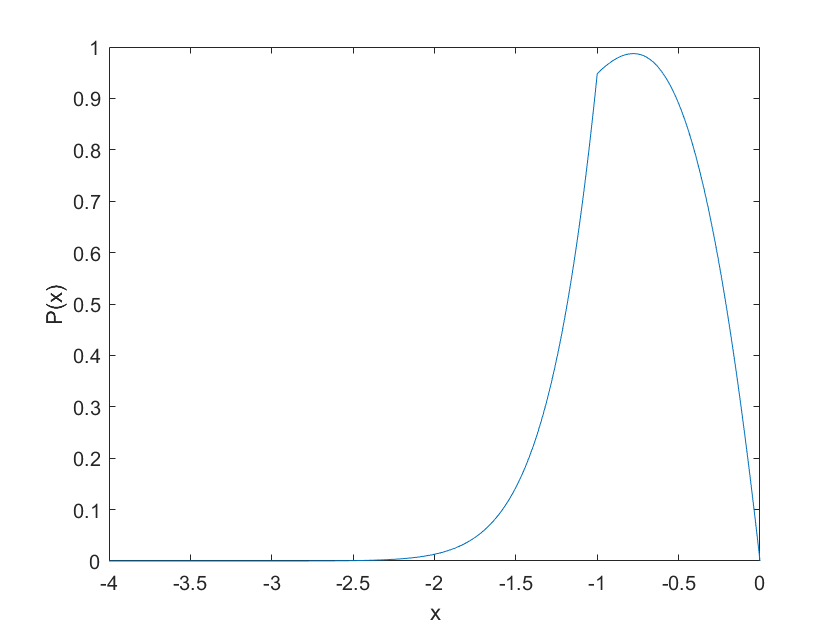}}
\\[0pt]
\subfloat[]{\includegraphics[width=0.5\textwidth]
{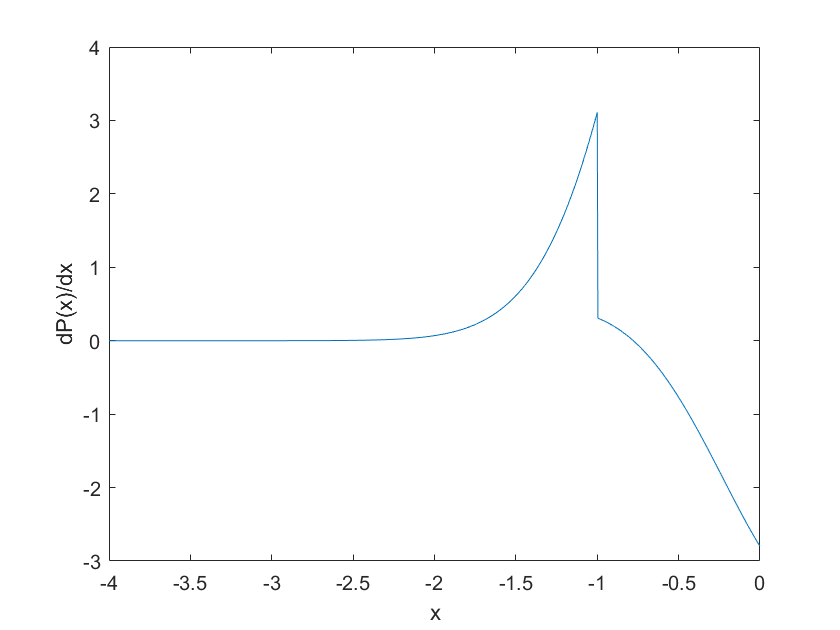}}
\end{center}
\par
\vspace{-10pt}
\caption{In Figures (a)-(b) we show the stationary distribution $p\left(
x\right) $ and its derivative $dp\left( x\right) /dx$ for the following
parameter values: $X_{0}=-1$, $m_{0}=0.5$, $m_{1}=0.1$. The corresponding
value of $\protect\lambda $, which is computed as part of the solution, is $%
1.4002$ .}
\label{Fig9}
\end{figure}

We show the stationary profile $p\left( x\right) $ and its derivative $%
dp\left( x\right) /dx$ in Figure \ref{Fig9}. As expected, $dp\left( x\right)
/dx$ jump down at $x=X_{0}$.

\begin{equation*}
\text{Figure \ref{Fig9} near here.}
\end{equation*}

\subsection{The nonstationary problem}

First, we use the following transformation of variables:%
\begin{equation*}
\begin{array}{c}
t=t,y=x-M\left( t\right) ,\ \ \ \ \ M\left( 0\right) =0,\ \ \ \frac{\partial 
}{\partial t}=\frac{\partial }{\partial t}-M^{\prime }\left( t\right) \frac{%
\partial }{\partial y},\ \ \ \ \ \frac{\partial }{\partial x}=\frac{\partial 
}{\partial y},%
\end{array}%
\end{equation*}%
and get the following IBVP:%
\begin{equation*}
\begin{array}{c}
\frac{\partial }{\partial t}p\left( t,y\right) =\frac{\partial }{\partial y}%
\left( \left( y+M^{\prime }\left( t\right) +M\left( t\right) -\mu \left(
t\right) \right) p\left( t,y\right) \right) +\frac{1}{2}\frac{\partial ^{2}}{%
\partial y^{2}}p\left( t,y\right) +\lambda \left( t\right) \delta
_{X_{0}-M\left( t\right) }\left( x\right) ,\ \ \infty <y\leq -M\left(
t\right) , \\ 
\\ 
p\left( 0,y\right) =\delta _{\xi }\left( y\right) ,\ \ \ p\left( t,-\infty
\right) =0,\ \ \ \ \ p\left( t,-M\left( t\right) \right) =0, \\ 
\\ 
\lambda \left( t\right) =-\frac{1}{2}\frac{\partial }{\partial y}p\left(
t,-M\left( t\right) \right) ,\ \ \ \ \ \mu \left( t\right)
=m_{0}+m_{1}\lambda \left( t\right) .%
\end{array}%
\end{equation*}%
Thus, by choosing $M$ in such a way that 
\begin{equation*}
\begin{array}{c}
M^{\prime }\left( t\right) +M\left( t\right) -\mu \left( t\right) =0,\ \ \ \
\ M\left( 0\right) =0,%
\end{array}%
\end{equation*}%
or, explicitly,%
\begin{equation*}
\begin{array}{c}
M\left( t\right) =\int_{0}^{t}e^{-\left( t-t^{\prime }\right) }\mu \left(
t^{\prime }\right) dt^{\prime },%
\end{array}%
\end{equation*}%
we get the IBVP\ for the standard Ornstein-Uhlenbeck process:%
\begin{equation*}
\begin{array}{c}
\frac{\partial }{\partial t}p\left( t,y\right) =\frac{\partial }{\partial y}%
\left( yp\left( t,y\right) \right) +\frac{1}{2}\frac{\partial ^{2}}{\partial
y^{2}}p\left( t,y\right) +\lambda \left( t\right) \delta _{X_{0}-M\left(
t\right) }\left( x\right) ,\ \ \ \infty <y\leq -M\left( t\right) , \\ 
\\ 
p\left( 0,y\right) =\delta _{\xi }\left( y\right) ,\ \ \ p\left( t,-\infty
\right) =0,\ \ \ \ \ p\left( t,-M\left( t\right) \right) =0, \\ 
\\ 
\lambda \left( t\right) =-\frac{1}{2}\frac{\partial }{\partial y}p\left(
t,-M\left( t\right) \right) ,\ \ \ \ \ \mu \left( t\right)
=m_{0}+m_{1}\lambda \left( t\right) .%
\end{array}%
\end{equation*}

As usual, we split $p\left( t,x\right) $ as follows:%
\begin{equation*}
\begin{array}{c}
p\left( t,x\right) =e^{t}H\left( \eta \left( t\right) ,e^{t}y-\xi \right)
+r\left( t,x\right) ,%
\end{array}%
\end{equation*}%
where the first term solves the governing equation and satisfies the
initial, but not the boundary conditions, while $r\left( t,x\right) $ solves
the following IBVP:%
\begin{equation*}
\begin{array}{c}
\frac{\partial }{\partial t}r\left( t,y\right) =\frac{\partial }{\partial y}%
\left( yr\left( t,y\right) \right) +\frac{1}{2}\frac{\partial ^{2}}{\partial
y^{2}}r\left( t,y\right) +\lambda \left( t\right) \delta _{X_{0}-M\left(
t\right) }\left( x\right) ,\ \ \ \infty <y\leq -M\left( t\right) , \\ 
\\ 
r\left( 0,y\right) =0,\ \ \ r\left( t,-\infty \right) =0,\ \ \ \ \ r\left(
t,-M\left( t\right) \right) =\chi _{0}\left( t\right) ,\  \\ 
\\ 
\lambda \left( t\right) =-\frac{1}{2}\frac{\partial }{\partial y}r\left(
t,-M\left( t\right) \right) +\chi _{1}\left( t\right) ,\ \ \ \ \ \mu \left(
t\right) =m_{0}+m_{1}\lambda \left( t\right) ,%
\end{array}%
\end{equation*}%
where%
\begin{equation*}
\begin{array}{c}
\chi _{0}\left( t\right) =-e^{t}H\left( \eta \left( t\right) ,e^{t}M\left(
t\right) +\xi \right) , \\ 
\\ 
\chi _{1}\left( t\right) =-\frac{e^{2t}\left( e^{t}M\left( t\right) +\xi
\right) }{2\eta \left( t\right) }H\left( \eta \left( t\right) ,e^{t}M\left(
t\right) +\xi \right) .%
\end{array}%
\end{equation*}%
We apply the familiar change of variables (\ref{Eq71}) and get the following
IBVP for $q\left( \tau ,\theta \right) =e^{-t}r\left( t,y\right) $: 
\begin{equation}
\begin{array}{c}
\frac{\partial }{\partial \tau }q\left( \tau ,\theta \right) =\frac{1}{2}%
\frac{\partial ^{2}}{\partial \theta ^{2}}q\left( t,\theta \right)
+\varkappa \left( \tau \right) \delta _{X_{0}-M\left( t\right) }\left(
x\right) ,\ \ \ \infty <\theta \leq \Gamma \left( \tau \right) , \\ 
\\ 
q\left( 0,\theta \right) =0,\ \ \ q\left( \tau ,-\infty \right) =0,\ \ \
q\left( \tau ,\Gamma \left( \tau \right) \right) =\varrho _{0}\left( \tau
\right) , \\ 
\\ 
\varkappa \left( \tau \right) =-\frac{\left( 1+2\tau \right) }{2}\frac{%
\partial }{\partial \theta }q\left( \tau ,\Gamma \left( \tau \right) \right)
+\varrho _{1}\left( \tau \right) ,%
\end{array}
\label{Eq113}
\end{equation}%
where%
\begin{equation*}
\begin{array}{c}
\Gamma _{0}\left( \tau \right) =\sqrt{1+2\tau }\left( X_{0}-M\left( \ln
\left( \sqrt{1+2\tau }\right) \right) \right) , \\ 
\\ 
\Gamma \left( \tau \right) =-\sqrt{1+2\tau }M\left( \ln \left( \sqrt{1+2\tau 
}\right) \right) , \\ 
\\ 
\varrho _{0}\left( \tau \right) =-H\left( \tau ,\Gamma \left( \tau \right)
-\xi \right) , \\ 
\\ 
\varrho _{1}\left( \tau \right) =\frac{\left( \Gamma \left( \tau \right)
-\xi \right) }{2\tau }H\left( \eta \left( t\right) ,\Gamma \left( \tau
\right) -\xi \right) .%
\end{array}%
\end{equation*}

Denoting $q\left( \tau ,\Gamma _{0}\left( \tau \right) \right) $ by $\nu
\left( \tau \right) $, we can split the IBVP (\ref{Eq113}) into two IBVPs:%
\begin{equation*}
\begin{array}{c}
\frac{\partial }{\partial \tau }q^{>}\left( \tau ,\theta \right) =\frac{1}{2}%
\frac{\partial ^{2}}{\partial \theta ^{2}}q^{>}\left( t,\theta \right) ,\ \
\ \Gamma _{0}\left( \tau \right) \leq \theta \leq \Gamma \left( \tau \right)
, \\ 
\\ 
q^{>}\left( 0,\theta \right) =0,\ \ \ q\left( \tau ,\Gamma _{0}\left( \tau
\right) \right) =\nu \left( \tau \right) ,\ \ \ \ \ q\left( \tau ,\Gamma
\left( \tau \right) \right) =\varrho _{0}\left( \tau \right) ,%
\end{array}%
\end{equation*}%
\begin{equation*}
\begin{array}{c}
\frac{\partial }{\partial \tau }q^{<}\left( \tau ,\theta \right) =\frac{1}{2}%
\frac{\partial ^{2}}{\partial \theta ^{2}}q^{<}\left( t,\theta \right) ,\ \
\ \infty <\theta \leq \Gamma _{0}\left( \tau \right) , \\ 
\\ 
q^{<}\left( 0,\theta \right) =0,\ \ \ q\left( \tau ,\theta \rightarrow
-\infty \right) \rightarrow 0,\ \ \ q\left( \tau ,\Gamma _{0}\left( \tau
\right) \right) =\nu \left( \tau \right) ,%
\end{array}%
\end{equation*}%
and a matching condition:%
\begin{equation*}
\frac{\partial }{\partial \theta }q^{<}\left( t,\Gamma _{0}\left( \tau
\right) \right) -\frac{\partial }{\partial \theta }q^{>}\left( t,\Gamma
_{0}\left( \tau \right) \right) =2\left( -\frac{\left( 1+2\tau \right) }{2}%
\frac{\partial }{\partial \theta }q^{>}\left( \tau ,\Gamma \left( \tau
\right) \right) +\varrho _{1}\left( \tau \right) \right) .
\end{equation*}%
We can now use results from Section \ref{Mathprel} to reduce these equations
to a very efficient (but highly nonlinear) system of Volterra integral
equations. Due to the lack of space, an analysis of the corresponding system
will be presented elsewhere.

\section{Conclusions}

In this document, we describe an analytical framework for solving several
relevant and exciting problems of financial engineering. We show that the
EMHP is a powerful tool for reducing partial differential equations to
integral equations of Volterra type. Due to their unique nature, these
equations are relatively easy to solve. In some cases, we can solve these
equations analytically by judiciously using the Laplace transform. In other
cases, we can solve them numerically by constricting highly accurate
numerical quadratures. We demonstrate that the EMHP has numerous
applications in mathematical finance and far beyond its confines.

\begin{acknowledgement}
Valuable discussions with our Investimizer colleagues Dr. Marsha Lipton, and
Dr. Marcos Lopez de Prado are gratefully acknowledged.
\end{acknowledgement}

\begin{acknowledgement}
The contents of this document were presented at a conference at the Hebrew
University of Jerusalem in December 2018. Exemplary efforts of the
organizers, Dr. David Gershon and Prof. Mathieu Rosenbaum, are much
appreciated.
\end{acknowledgement}

\begin{acknowledgement}
Some of the ideas described in this document were developed jointly with Dr.
Vadim Kaushansky and Prof. Christoph Reisinger.
\end{acknowledgement}

\end{document}